\documentclass[journal]{IEEEtran}
 
\usepackage{cite}

\usepackage[pdftex]{graphicx}
\usepackage{subfigure}
\usepackage{float}
\usepackage[colorlinks]{hyperref}
\graphicspath{{../pdf/}{../jpeg/}}
\DeclareGraphicsExtensions{.pdf,.jpeg,.png}

\usepackage{tabu}
\usepackage{multirow}
\usepackage{colortbl}
\usepackage{booktabs}
\usepackage{amsmath}
\usepackage{amsfonts}

\usepackage[dvipsnames, svgnames, x11names]{xcolor}

\begin{document}
	
	\title{TransBTSV2: Towards Better and More Efficient Volumetric Segmentation of Medical Images}
	
	\author{Jiangyun Li,
		Wenxuan Wang, 
		Chen Chen,~\IEEEmembership{Member,~IEEE,}
		Tianxiang Zhang,
		Sen Zha,
		Jing Wang,
		Hong Yu
		\thanks{J. Li, W. Wang, T. Zhang, S. Zha, J. Wang and H. Yu are with the School of Automation and Electrical Engineering, University of Science and Technology Beijing,
			Beijing 100083, China. (e-mail: leejy@ustb.edu.cn, s20200579@xs.ustb.edu.cn, txzhang@ustb.edu.cn, g20198675@xs.ustb.edu.cn, m202120718@xs.ustb.edu.cn,
			g20198754@xs.
			\newline ustb.edu.cn)}
		\thanks{C. Chen is with the Center for Research in Computer Vision, University of Central Florida, Orlando, FL 32816 USA. (e-mail: chen.chen@crcv.ucf.edu)}
		\thanks{Corresponding author: Jiangyun Li.}
	}
	
	
	\maketitle
	
	\begin{abstract}
		Transformer, benefiting from global (long-range) information modeling using self-attention mechanism, has been successful in natural language processing and computer vision recently. Convolutional Neural Networks, capable of capturing local features, are difficult to model explicit long-distance dependencies from global feature space.
        However, both local and global features are crucial for dense prediction tasks, especially for 3D medical image segmentation.
        In this paper, we present the further attempt to exploit Transformer in 3D CNN for 3D medical image volumetric segmentation and propose a novel network named TransBTSV2 based on the encoder-decoder structure. Different from TransBTS \cite{wang2021transbts}, the proposed TransBTSV2 is not limited to brain tumor segmentation (BTS) but focuses on general medical image segmentation, providing a stronger and more efficient 3D baseline for volumetric segmentation of medical images. 
        As a hybrid CNN-Transformer architecture, TransBTSV2 can achieve accurate segmentation of medical images without any pre-training, possessing the strong inductive bias as CNNs and powerful global context modeling ability as Transformer. 
        With the proposed insight to redesign the internal structure of Transformer block and the introduced Deformable Bottleneck Module to capture shape-aware local details, a highly efficient architecture is achieved with superior performance. Extensive experimental results on four medical image datasets (BraTS 2019, BraTS 2020, LiTS 2017 and KiTS 2019) demonstrate that TransBTSV2 achieves comparable or better results compared to the state-of-the-art methods for the segmentation of brain tumor, liver tumor as well as kidney tumor. Code will be publicly available at \url{https://github.com/Wenxuan-1119/TransBTS}.
	\end{abstract}
	

	\IEEEpeerreviewmaketitle

	\section{Introduction}
	\label{sec:introduction}

\begin{figure}[!tp]
        \centering
        \rotatebox{90}
        {
        \begin{tabu} to 0.6\linewidth{X[1.0c] X[1.0c] } 
            \scriptsize{Ground Truth} &  \scriptsize{Images}  \\
        \end{tabu}
        }
        \includegraphics[width=0.45\textwidth]{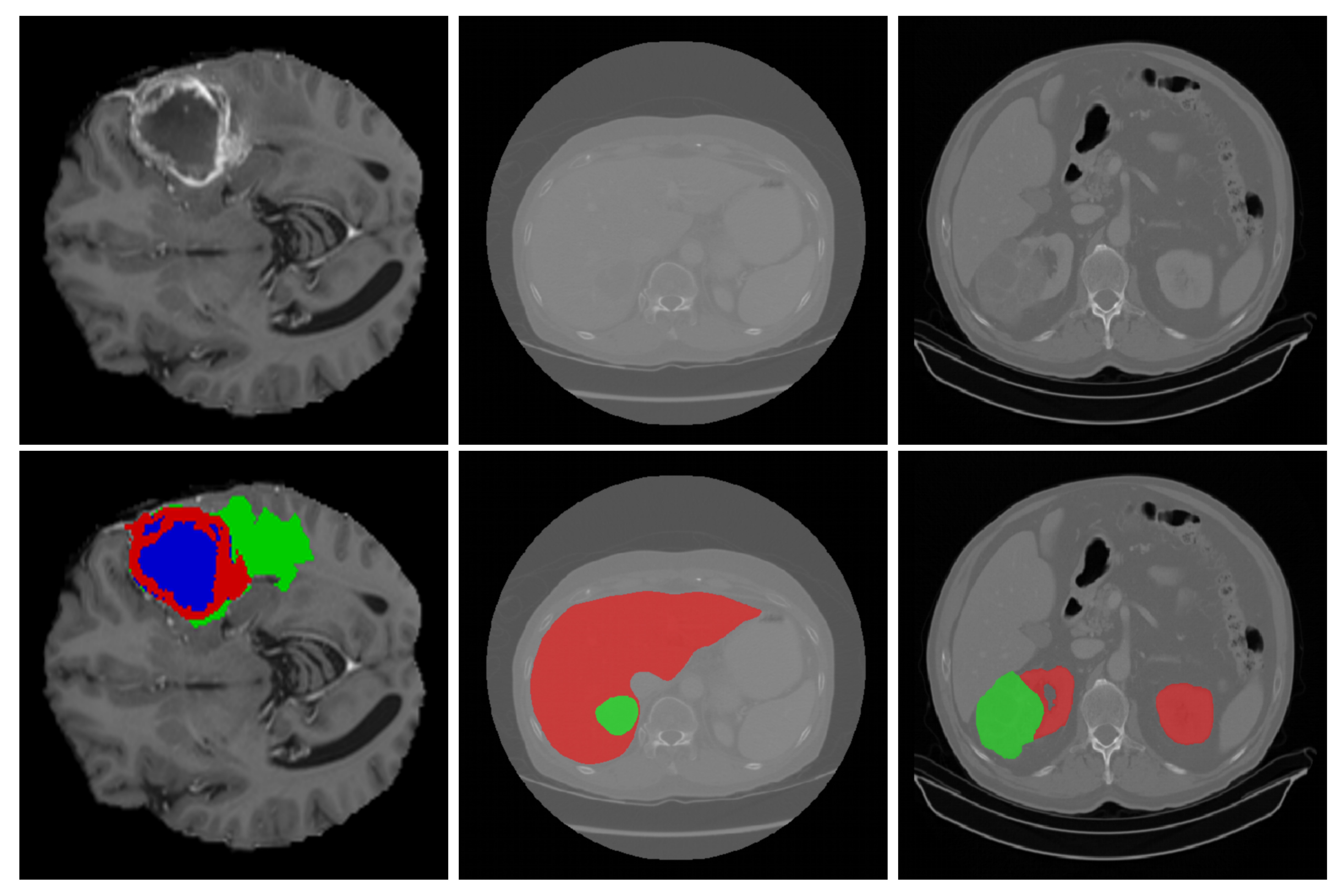}
        \begin{tabu} to 1.03\linewidth{X[1.06c] X[0.66c] X[1.42c]} 
            \scriptsize{Brain Tumor} &  \scriptsize{Liver Tumor} &  \scriptsize{Kidney Tumor} \\
        \end{tabu}
        \caption{Examples of medical images with the corresponding semantic segmentation annotations.}
        \label{fig1}
    \end{figure}

    \IEEEPARstart{A}{s} one of the most common clinical diseases, cancer causes numerous deaths every year. The precise measurements from medical images can assist doctors in making accurate diagnosis and further treatment planning. As illustrated in Fig. \ref{fig1}, medical image segmentation aims to identify tumors and delineate different sub-regions of organs from background, by assigning a predefined class label to each pixel in medical images, such as Magnetic Resonance Imaging (MRI) \cite{huo2017robust} and Computerized Tomography (CT) \cite{nguyen2015segmentation}. Traditionally, the lesion regions are mainly delineated by clinicians, heavily relying on clinical experiences, which is time-consuming and prone to errors. 
    Therefore, for improving the accuracy and efficiency of clinical diagnosis, it is of vital importance to promote the development of automatic medical image segmentation.

    Different from natural images, some unique characteristics of medical images make the automatic segmentation task very challenging. First, the shapes and textures of different tumors vary greatly, thus it is difficult to find the commonalities between them by direct matching. In addition, some lesions do not have clear boundaries so that accurately separating the contours of these tumors is really difficult. Furthermore, differences of tumor structure in size, extension and location also prevent segmentation algorithms from using strong prior knowledge. Therefore, it is very challenging to automatically and accurately segment medical images.
    
    To tackle the aforementioned problems of medical image segmentation, many approaches have been proposed, including statistical shape models, contour-based approaches and machine learning-based methods \cite{yu2006medical, tsai2003shape, held1997markov}. However, these methods heavily rely on hand-crafted features and have limited feature representation capability. Recently, Convolutional Neural Networks (CNNs) have dominated visual modeling in a wide range of computer vision tasks.
    Since local features are essential for the segmentation of boundaries and lesions in medical images, CNNs are quickly adopted for medical image segmentation. In particular, U-Net \cite{unet}
    adopts a symmetric encoder-decoder structure with skip-connections to refine detail preservation. With the skip-connections and successive upsampling operations in decoder, U-Net can realize the fusion of multi-level features and recover more accurate segmentation edges, becoming the mainstream architecture for medical image segmentation. In the following research, U-Net variants such as U-Net++ \cite{unet++} and Res-UNet \cite{zhang2018road} further improve the segmentation performance. However, although CNN-based methods have prominent representation capabilities, it is difficult for CNNs to model explicit \textbf{long-range} dependencies due to limited receptive fields of convolution kernels. This intrinsic limitation of convolution operations prevents CNNs from learning global semantic information, which is critical for dense prediction tasks like segmentation.

    Inspired by attention mechanism \cite{bahdanau2014neural} in natural language processing (NLP), existing research overcome the inherent inability to model long-distance dependencies by fusing the attention mechanism with CNN models. The key idea of the attention mechanism is to model the long-range relationship between feature representations and update feature maps based on the attention weights. Different from convolution, a single attention layer is capable of capturing global semantic information from the whole feature space.
    For example, non-local neural networks \cite{wang2018non} aim to capture the long-distance dependencies in feature maps based on the self-attention mechanism, but suffering from the high memory and computational cost. Since modeling explicit long-range dependencies is crucial for precise segmentation of organs and tumors in medical images, attention mechanism is also widely adopted in medical image segmentation \cite{oktay2018attention,schlemper2019attention}, obtaining remarkable results.

    Based on self-attention mechanism, Transformer \cite{vaswani2017attention}, designed to model long-range dependencies in sequence-to-sequence tasks, is able to capture the relations between arbitrary positions in the sequence. Compared to attention mechanism, Transformer is more powerful in modeling global context and possess stronger capability of feature representation attributed to its unique architecture, which is more suitable for overcoming the difficulties of medical image segmentation. 
    Lately, Transformer-based frameworks have reached state-of-the-art performance on various computer vision tasks. For example, Vision Transformer (ViT) architecture \cite{vit,deit} has achieved outstanding performance for image classification via pre-training on large-scale datasets. Afterwards, many Transformer-based architectures such as \cite{zheng2021rethinking,wang2021pyramid,liu2021swin} have been explored for dense prediction tasks.

    Although Transformer-based methods have achieved promising results on various computer vision tasks, training and deploying Transformer is still a great challenge. On the one hand, the original self-attention mechanism in vanilla Transformer leads to $O(n^{2})$ time and space complexity with respect to sequence length. Furthermore, related research usually employ multiple Transformer blocks, resulting in unbearable computational cost and memory footprint, especially when the sequence length is extremely long. To address this issue, many recent research \cite{kitaev2020reformer, beltagy2020longformer, zaheer2020big} have focused on reducing the computational complexity of vanilla Transformer. On the other hand, distinct from CNNs, the performance of Transformer heavily depends on the scale of datasets. As Transformer do not possess strong inductive bias as convolution operations, the performance of Transformer is usually inferior to CNNs when the training data is scarce. To alleviate this problem, a number of state-of-the-art methods turn to pre-training on large-scale datasets. However, medical image datasets are generally short of available training samples, making the pre-training of Transformer on medical images impractical.

    In this paper, we present the further attempt towards better and more efficient volumetric segmentation of medical images. 
    Specifically, since spatial and slice information are both crucial for volumetric segmentation of medical images, we correspondingly exploit Transformer in 3D CNN to effectively capture both local and global volumetric spatial features. 
    To fully exploit the merits of both CNN and Transformer, the network encoder first utilizes 3D CNN to extract the volumetric spatial features and downsample the input 3D images at the same time, effectively capturing the local 3D context information and resulting in compact volumetric feature maps. Then each volume is reshaped into a vector (i.e. token) and fed into Transformer for global feature modeling. Finally, the 3D CNN decoder takes the feature embeddings from Transformer as well as the low-level feature from skip-connections, and performs progressive upsamplings to predict the full resolution segmentation maps. As a hybrid CNN-Transformer architecture, our TransBTSV2 can perform accurate segmentation of medical images without any pre-training.

    Since the conventional design of Transformer-based methods is to repeatedly stack Transformer blocks along model depth, the introduction of Transformer brings considerable computational overhead. To address this dilemma, inspired by the inverted design in MobileNetV2 \cite{sandler2018mobilenetv2}, we propose a novel insight to pursue wider instead of deeper Transformer architecture to improve model efficiency. Following this simple yet effective design scheme, we gradually expand the model width twice in the whole pipeline to achieve an inverted bottleneck alike architecture, which leads to an impressive decrease in model complexity ($\textbf{53.62\%}$ reduction in parameters and $\textbf{27.75\%}$ reduction in FLOPs) compared with our original TransBTS \cite{wang2021transbts}. 
    \textit{With only a single Transformer block}, superior performance is achieved by our highly efficient TransBTSV2.

    In addition, the irregular-shape lesions also bring great challenges to medical image segmentation. Due to the fixed geometric structures of CNN basic modules (e.g. convolution and pooling), CNNs are inherently limited to model irregular-shape deformation of lesion regions. In U-Net architecture, feature maps from encoder are more sensitive to the geometry information and essential to the recognition of target areas. To this end, we propose an effective and efficient Deformable Bottleneck Module (DBM) at the position of each skip-connection that can learn volumetric spatial offsets from the encoder features and adapt to various transformations of segmentation targets with nearly negligible extra computational costs. Thus, with the benefit of the introduced deformable convolution \cite{dai2017deformable} in our proposed DBM, the feature maps output from DBMs capture more shape-aware local details and the proposed network can generate finer segmentation results of lesions.

    Comprehensive experiments are conducted on four benchmark datasets for medical volumetric segmentation to shed light on architecture engineering of incorporating Transformer in 3D CNN. As a hybrid CNN-Transformer architecture, our TransBTSV2 can not only inherits the great generalization property but also enjoys the superior long-range dependencies modeling capability of Transformer, unleashing the power of both architectures. It is worth noting that TransBTSV2 is a clean and general 3D network (with basic 3D convolutional layers and Transformer blocks) without any complex structures and add-ons. Any effective techniques such as multi-scale feature fusion can be easily plugged into TransBTSV2 to boost the performance. 
    
    This work is an extension of our MICCAI version \cite{wang2021transbts}, which presents the first attempt yo exploit Transformer in 3D CNN for 3D MRI brain tumor segmentation. In this paper, we make the following important extensions:

    \begin{itemize}
        \item We deviate from the design pattern of traditional Transformer-based methods and propose a novel insight to redesign the internal structure of Transformer block, pursuing a shallower but wider architecture rather than the conventional deeper but narrower architecture. The model complexity is greatly reduced while better performance is also achieved for medical image segmentation.
    
        \item We propose a general framework TransBTSV2 for volumetric segmentation of medical images and introduce a Deformable Bottleneck Module (DBM) to capture more shape-aware feature representations with nearly negligible extra computational overhead. By employing the proposed DBM, our method can generate finer segmentation results of lesion areas.
    
        \item Compared with our MICCAI version \cite{wang2021transbts}, we present more results on two popular CT image datasets (LiTS 2017 and KiTS 2019) to evaluate the generalization ability of our proposed TransBTSV2. 
        Extensive experimental results on four benchmark datasets for medical volumetric segmentation demonstrate that TransBTSV2 reaches competitive or better performance than previous state-of-the-art methods. 
        
        \item We provide a comprehensive comparison between TransBTS \cite{wang2021transbts} and the proposed TransBTSV2 in terms of network optimization, feature representation and confidence distribution of model prediction to fully validate the superiority of TransBTSV2.
    \end{itemize}
    
    The remainder of this paper is organized as follows. Sec.~\ref{sec:relatedwork} briefly reviews the related works, and Sec.~\ref{sec:methodology} provide the details of the proposed TransBTSV2. Then, the experimental results and corresponding analysis are presented in Sec.~\ref{sec:experimentsandresults}, which is followed by the conclusion in Sec.~\ref{sec:conclusion}.

	\section{Related Work}
	\label{sec:relatedwork}
	
    We present a brief overview of the related works from two aspects. The first aspect primarily introduces U-Net and its variants on medical image segmentation, while the second aspect reviews the Transformer-based approaches for various vision tasks, especially medical image segmentation.
    
    \subsection{U-Net and U-Net Variants}
    
    Due to the difficulties of medical image segmentation mentioned above, the multi-scale feature fusion and fine-grained local details modeling are both indispensable for an advanced network. 
    To preserve detail information of medical images as much as possible, U-Net \cite{unet} adopts a symmetric encoder-decoder architecture with skip connections to gradually restore the downsampled feature maps to original size, thereby achieving pixel-level dense prediction of medical images. In the following research, U-Net based variants attract a lot of attention and are further applied in medical image segmentation. To fully utilize the local information among continuous slices which is also critical for volumetric segmentation, Cicek et al. \cite{3dunet} generalize the U-Net from 2D to 3D by implementing 3D operations, such as 3D convolution and 3D max pooling, for better medical volumetric segmentation. To model long-range distance dependencies, Ozan Oktay \cite{oktay2018attention} proposes a novel attention gate for medical imaging that automatically learns to focus on target structures of varying shapes and sizes. Based on the residual structure of ResNet \cite{he2016deep}, Residual U-Nets \cite{fang2019rca, xiao2018weighted} are also proposed to overcome the difficulty in training deep neural networks. In addition, there are works \cite{unet++, huang2020unet} aim to enrich the information interaction of feature maps between different levels by improving skip connections.

    \subsection{Transformer-based Approaches}
    \subsubsection{Transformer for Various Vision Tasks}
    
    With the help of pre-training on large-scale datasets, Transformer-based frameworks have reached state-of-the-art performance on various vision tasks recently. 
    Vision Transformer (ViT) \cite{vit} splits the image into fixed-size patches and models the correlations between these patches as sequences, achieving satisfactory results. DeiT \cite{deit} introduces a knowledge distillation method for training ViT to further improve its performance. DETR \cite{detr} treats object detection as a set prediction task using Transformer, which simplifies the object detection process by abandoning the steps of generating anchors and non-maximum suppression. 
    Rethinking semantic segmentation from a sequence-to-sequence perspective, SETR \cite{zheng2021rethinking} leverages Transformer as the encoder for global feature extraction and achieves superior performance with large-scale pre-training. 
    To construct efficient architecture for segmentation task, SegFormer \cite{xie2021segformer} combines a novel hierarchically structured Transformer encoder with a lightweight multi-layer perceptron (MLP) decoder. Moreover, to further achieve efficient architecture with less redundancy, Swin Transformer \cite{liu2021swin} gets rid of the global pairwise self-attention mechanism in ViT and proposes a new Transformer backbone based on the proposed local-window self-attention mechanism and shifted window scheme, reaching state-of-the-art results on various benchmarks.

    \subsubsection{Transformer for Medical Image Segmentation}
    
    To explicitly model long-distance dependencies of volumes in medical images, recent research focus on introducing Transformer to medical image segmentation task. 
    TransUNet \cite{chen2021transunet} employs ViT as encoder with large-scale pre-training for medical image segmentation. 
    Then, Swin-Unet \cite{cao2021swin} uses hierarchical Swin
    Transformer with shifted windows as the encoder to extract context features and adopts a symmetric Swin Transformer-based decoder as the decoder.
    MedT \cite{valanarasu2021medical} proposes a Gated Axial-Attention model which applies an additional control mechanism in the self-attention module to extend existing architectures. 
    FTN \cite{he2022fully} leverages sliding window tokenization to construct hierarchical features and a Transformer decoder to aggregate hierachical feafures, which is fully relied on Transformer for skin lesion analysis.
    FAT-Net \cite{wu2022fat} adopts feature adaptive Transformers for automated skin lesion segmentation and replace the traditional single branch encoder with a dual branch encoder of both CNNs and Transformers.
    In addition, by employing the proposed relation transformer block (RTB) and global transformer block (GTB), RTNet \cite{huang2022rtnet} can achieve the accurate segmentation of diabetic retinopathy lesions.
    Different from these previously proposed methods, we focus on investigating how to effectively and efficiently incorporate Transformer in popular 3D CNNs to capture volumetric spatial information, unleashing the potential of both networks. Rather than just surpassing the SOTA results with complex designs, our TransBTSV2 is a highly efficient framework with clean structure and superior performance.

	\section{Methodology}
    \label{sec:methodology}
    In this section, we first briefly introduce the overall architecture of our TransBTSV2 and provide the details of TransBTSV2 in terms of network encoder and decoder. Then, we present the novel insight to redesign Transformer block to achieve the highly efficient architecture.
    Finally, we introduce the Deformable Bottleneck Module in TransBTSV2 for capturing more shape-aware feature representations.
    
    \begin{figure*}
        \centering
        \includegraphics[width=1.0\textwidth]{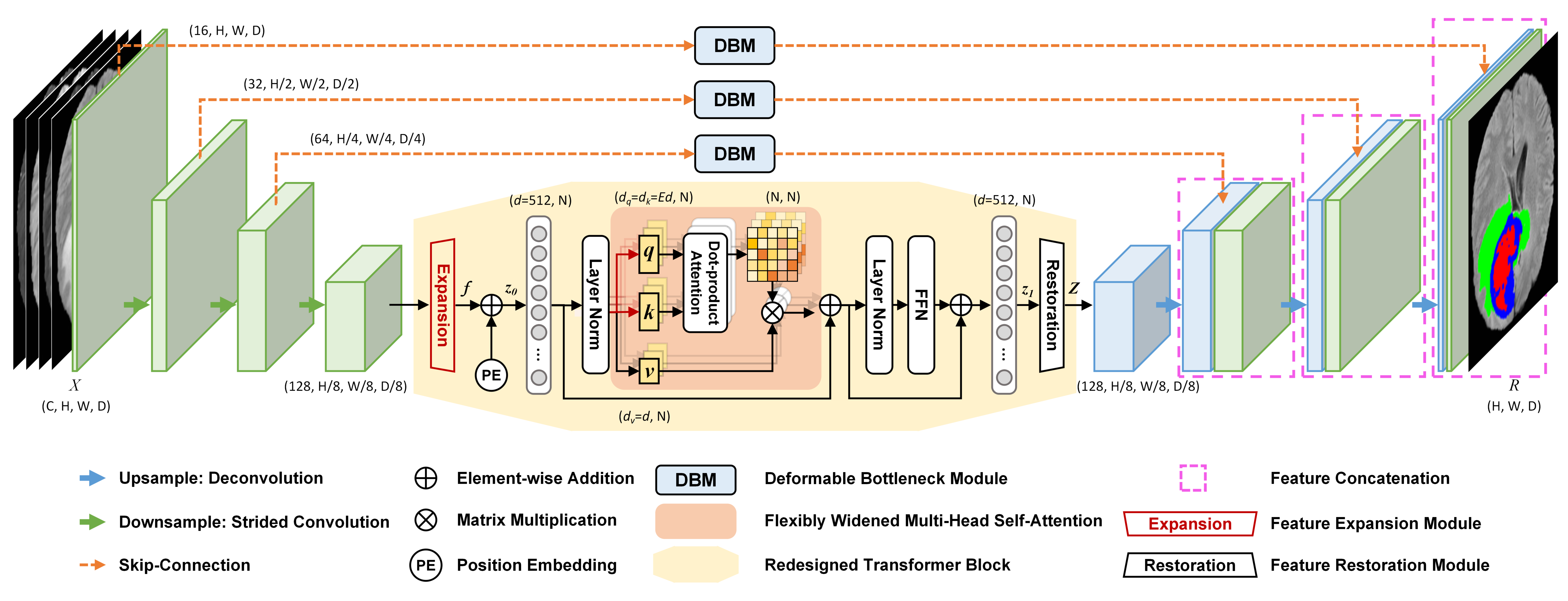}
        \caption{The illustration of the proposed TransBTSV2 for automatic medical image segmentation. We use modified 3D CNN encoder to capture local information and leverage the Transformer encoder to model long-distance dependencies from the global view. Upsampling and convolutional layers are stacked to gradually produce high-resolution segmentation results. 
        We expand the model width totally twice (as highlighted with \textcolor{red}{red} color in the figure) in the whole architecture. With an inverted bottleneck alike Transformer architecture (as highlighted by the \textcolor{Wheat}{yellow} shadow in the figure) and the proposed DBM, TransBTSV2 is a highly efficient architecture with superior performance and low model complexity.}
        \label{fig2}
    \end{figure*}
    
    \subsection{CNN-Transformer Hybrid Architecture}
    \subsubsection{Overall Architecture of TransBTSV2}
    An overview of the proposed TransBTSV2 is presented in Fig. \ref{fig2}. 
    Given an input medical image $X \in \mathbb{R}^{C \times H \times W \times D}$ with a spatial resolution of $H \times W$, depth dimension of $D$ (\# of slices) and $C$ channels (\# of modalities), we first utilize modified 3D CNN to efficiently generate compact feature maps capturing volumetric spatial features, and then leverage the redesigned Transformer encoder to model the long-distance dependencies in a global space. After that, we repeatedly apply the upsampling and convolutional layers to gradually produce a high-resolution segmentation result. 
    
    \subsubsection{Network Encoder}
    
    Due to the computational complexity of Transformer is quadratic with respect to the number of tokens (i.e. sequence length), directly flattening input image to sequence as Transformer input is impractical. Therefore, ViT \cite{vit} splits an image into fixed-size ($16 \times 16$) patches and then reshapes each patch into a token. Following ViT, the straightforward tokenization is splitting images into 3D patches for volumetric data. However, this simple strategy makes Transformer unable to model the image \textit{local context information across spatial and depth dimensions} for volumetric segmentation. 
    To address this issue, our solution is to employ a stack of convolutional layers with downsampling (strided convolution with stride=2) to gradually encode input images into low-resolution/high-level feature representations $F \in \mathbb{R}^{K \times \frac{H}{8} \times \frac{W}{8} \times \frac{D}{8}}$ ($K=128$), which is 1/8 of input dimensions of $H, W$ and $D$ ({overall stride (OS)=8}). In this way, rich local 3D context features are effectively embedded in $F$ with a relatively small computation budget in the early stage.
    Then, $F$ is fed into the Transformer encoder to further learn long-range correlations with global receptive field.

    \noindent \textbf{Feature Embedding of Transformer Encoder.}
    Given the feature maps $F$, a feature expansion module is designed to realize the feature embedding for the following Transformer encoder. Concretely, a $3 \times 3 \times 3$ convolutional layer is firstly used to increase the channel dimension from $K=128$ to $d=512$, ensuring comprehensive representation of each volume and expanding the internal width of the following Transformer blocks simultaneously. As the Transformer block expects a sequence as input, we collapse the spatial and depth dimensions into one dimension, resulting in a $d \times N$ $(N=\frac{H}{8} \times \frac{W}{8} \times \frac{D}{8})$ feature map $f$ (i.e. $N$ $d$-dimensional tokens). To encode the location information that is necessary for segmentation task, we introduce the learnable position encodings and fuse them with the feature map $f$ by direct addition, creating the feature embeddings as follows:
    \begin{equation}
        z_{0}=f+PE=W \cdot F+PE
        \label{equation1}
    \end{equation}
    where $W$ is the feature expansion module, $PE \in \mathbb{R}^{d \times N}$ denotes the position encodings, and $z_{0} \in \mathbb{R}^{d \times N}$ refers to the feature embeddings.
    
    \noindent \textbf{Transformer Blocks.}
    The Transformer encoder is composed of $L$ redesigned Transformer blocks, and each of them has a modified architecture, consisting of a flexibly widened multi-head self-attention (FW-MHSA) block and a feed-forward Network (FFN). The output of the $\ell$-th ($\ell \in [1,2,...,L]$) Transformer block can be calculated by:
    \begin{equation}
        z_{\ell-1}^{'}=FW\textbf{-}MHSA(LN(z_{\ell-1}))+z_{\ell-1}
    \end{equation}
    \begin{equation}
        z_{\ell}=FFN(LN(z_{\ell-1}^{'}))+z_{\ell-1}^{'}
    \end{equation}
    where $LN(*)$ is the layer normalization and $z_{\ell}$ is the output of $\ell$-th Transformer block. 
    
    \subsubsection{Network Decoder}
    \label{decoder}
    In order to generate full-resolution segmentation results as original 3D image space ($H \times W \times D$), we introduce 3D CNN decoder to perform feature upsamplings and pixel-level segmentation (see the right part of Fig.~\ref{fig2}). 
    
    \noindent \textbf{Feature Restoration.}
    To fit the input dimension of 3D CNN decoder, we design a feature restoration module to project the sequence data back to a standard 4D feature map. Specifically, the output sequence of Transformer $z_{L} \in \mathbb{R}^{d \times N}$ is initially reshaped to $d \times \frac{H}{8} \times \frac{W}{8} \times \frac{D}{8}$. In order to reduce the computational complexity of decoder, two convolutional layers are employed to reduce the channel dimension from $d$ to $K$. Through these operations, the feature maps $Z \in \mathbb{R}^{K \times \frac{H}{8} \times \frac{W}{8} \times \frac{D}{8}}$ is obtained, with the same dimension as $F$ in the feature encoding part.

    \noindent \textbf{Progressive Feature Upsampling.}
    After the feature mapping, cascaded upsampling operations and convolutional layers are applied to $Z$ to gradually recover a full resolution segmentation result $ R \in \mathbb{R}^{H \times W \times D} $. Moreover, skip-connections are employed to fuse the encoder features with the decoder counterparts by concatenation for finer segmentation masks with richer spatial details.

    \subsection{Deviating from the Design Pattern of  Traditional Transformer-based Methods: Prefer Wider to Deeper}
    \label{Wider}
    
    To cope with the computational overhead brought by Transformer, inspired by inverted bottleneck blocks in MobileNetV2 \cite{sandler2018mobilenetv2}, a novel insight of designing Transformer is proposed in this work. In our original TransBTS \cite{wang2021transbts}, the number of Transformer blocks is $L$ = 4 and the Transformer part accounts for 70.81$\%$ of model parameters. Thus, shrinking the model size of Transformer is essential for reducing the overall model complexity. Recently, the mainstream design for Transformer-based methods is to pursue a deeper but narrower hierarchical architecture. However, in this paper, we redesign the Transformer architecture to go wider (hidden dimension of feature vectors) instead of deeper (number of Transformer blocks). First, the number of Transformer blocks $L$ is reduced from 4 to 1 in order to cut down the model size, leading to an impressive decrease in model complexity ($\downarrow\textbf{54.11\%}$ in parameters and $\downarrow\textbf{37.54\%}$ in FLOPs). However, the remarkable reduction in parameters usually leads to worse feature representations and an inevitable decline in model performance. Therefore, to restore the modeling capability of our model, the internal width of Transformer is gradually expanded twice (as highlighted with red color in Fig.~\ref{fig2}). Specifically, we utilize a feature expansion module to enlarge Transformer width at the first time. Then, with an expansion ratio $E$, the hidden dimension of $q$ and $k$ are expanded to $d_m$ (i.e. $d_m$ = $Ed$) to further increase the Transformer width. $v$ remains unchanged (i.e. $d$ = 512) to keep the dimensions of input and output consistent. 
    
    As described above, a single scaled dot-product self-attention in the FW-MHSA block can be formulated as:
    \begin{equation}
    q\uparrow = W_q \cdot x_{i}
    \end{equation}
    \begin{equation}
    [k\uparrow; v] = [W_k; W_v] \cdot LN(DWConv(x_{i}))
    \end{equation}
    \begin{equation}
    W_{atten} = softmax(q \cdot k^\top / \sqrt{d})
    \end{equation}
    \begin{equation}
    x_{o} = W_{atten} \cdot v
    \end{equation}
    where $W_q \in{\mathbb{R}^{d \times d_m}}$, $W_k \in{\mathbb{R}^{d \times d_m}}$ and $W_v \in{\mathbb{R}^{d \times d}}$ are transformation matrices, transferring the input feature embeddings $x_{i} \in{\mathbb{R}^{N \times d}}$ to different dimensional vectors $q\uparrow \in{\mathbb{R}^{N \times d_m}}$, $k\uparrow \in{\mathbb{R}^{N \times d_m}}$, and $v \in{\mathbb{R}^{N \times d}}$ respectively. In addition, $W_{atten} \in{\mathbb{R}^{N \times N}}$ denotes the attention weights and $x_{o} \in{\mathbb{R}^{N \times d}}$ is the output feature embeddings. $DWConv$ is the 3D depth-wise convolution layer, which is introduced to bring the local inductive bias into the modified Transformer architecture and further control the computational complexity.
    
    In this way, the Transformer block in our model achieves an inverted bottleneck alike architecture (as delineated by the yellow shaded Transformer block in Fig.~\ref{fig2}) with greatly reduced model complexity, leading to the proposed highly efficient architecture. Overall, pursing the inverted bottleneck alike architecture of Transformer has three-fold advantages. First, the model can adaptively learn how to expand the low-dimensional feature vectors to high dimension space and compress the high-dimensional feature vectors back into low dimension, making the whole network excellent in scalability and robustness. Second, instead of using low-dimensional vectors to perform matrix operations, computing the multi-head self-attention with the expanded high-dimensional vectors can greatly improve feature representations of the whole architecture with almost negligible extra computational costs, thus further improving the model performance in a highly efficient manner. Lastly, compared with deeper and narrower architecture, the design of the wider and shallower counterpart allows for more parallel processing, easier optimization and greatly reduced latency.

    \begin{figure}[!t]
        \centering
        \includegraphics[width=0.5\textwidth]{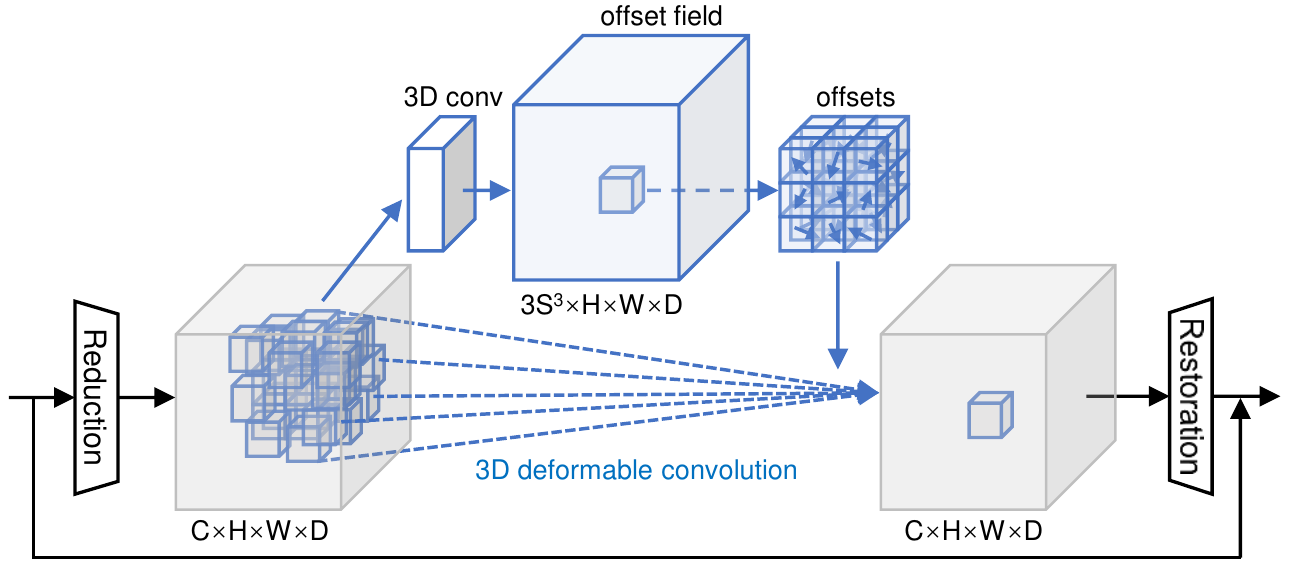}
        \caption{The illustration of the proposed Deformable Bottleneck Module. S refers to the kernel size of the 3D deformable convolution.}
        \label{fig3}
    \end{figure}
    
    \subsection{Deformable Bottleneck Module}
    \label{Deformable}

    As aforementioned in Sec.~\ref{sec:introduction}, since the lesion areas in medical images are commonly in irregular shapes with great varieties, it is very challenging to recognize the lesion areas precisely. To achieve better performance in medical image segmentation, it is necessary for the network to capture fine-grained and shape-aware local details.
    Due to the fixed geometric structures of CNN basic modules, CNNs are inherently limited to model irregular-shape deformation.
    To solve this problem, our proposed Deformable Bottleneck Module is designed to further capture shape-aware features from irregular-shape lesion regions. The details of Deformable Bottleneck Module are presented in Fig. \ref{fig3}. As the feature maps from encoder are more sensitive to geometry information and are essential to the recognition of target areas, the proposed DBMs are plugged right into each skip-connection. Each DBM consists of two separately placed convolutional blocks, 3D deformable convolutional layer\cite{dai2017deformable} and traditional residual connection for better optimization. In order to minimize the amount of computational overhead brought by the proposed DBM and realize the more reasonable resource allocation along model width, the two convolutional blocks (i.e. Reduction and Restoration layer illustrated in Fig. \ref{fig3}) are deployed at both ends of our DBM to reduce and restore channel dimensions, while the integrated 3D deformable convolution \cite{dai2017deformable} is utilized to capture shape-aware local information along volumetric spatial dimension. 
    
    As illustrated in Fig. \ref{fig3}, the learned volumetric spatial offsets are obtained by a 3D convolutional layer over the input feature maps. The volumetric spatial resolution of the learned offsets is the same as the input feature maps, and the channel dimension $3S^{3}$ corresponds to $S^{3}$ 3D offsets, where $S$ denotes the kernel size of the 3D deformable convolution. Each position of the 3D deformable convolution kernel corresponds to a volumetric spatial offset so that the sampling can be adapted to the irregular and offset locations instead of following the regular grid. Different from the 2D offsets learned by conventional 2D deformable convolution \cite{dai2017deformable} that only focus on spatial dimension, the volumetric spatial offsets efficiently acquired by our proposed DBM can help the sampling grid make adjustments at both spatial and slice dimension, further enabling our TransBTSV2 to address the great challenges brought by the irregular shapes of lesion areas.
    To be specific, the traditional convolution operation with the regular grid can be represented as:
    \begin{equation}
    y(p_0) = \sum_{p_n \in \Re} \omega(p_n) \cdot x(p_0 + p_n)
    \end{equation}
    where $\Re$ is the regular sampling grid, $x$ is the input feature maps, $y$ is the output feature maps and $\omega$ is the weights of convolution operation. Moreover, $p_0$ and $p_n$ denotes the voxel of the output feature maps and the locations in the enumerated convolution kernel respectively. In contrast, the modulated deformable convolution can be calculated as:
    \begin{equation}
    y(p_0) = \sum_{p_n \in \Re} \omega(p_n) \cdot x(p_0 + p_n + \bigtriangleup p_n)
    \end{equation}
    where $\bigtriangleup p_n$ represents the learned volumetric spatial offsets as presented in Fig. \ref{fig3}. The offset $\bigtriangleup p_n$ is added to the original sampling position so that the free deformation is realized by the irregular offset locations $p_n + \bigtriangleup p_n$.
    In this way, DBMs at each skip-connection can both effectively and efficiently learn volumetric spatial offsets from the encoder features and adapt to various transformations of segmentation targets with nearly negligible introduced computational overhead. Thus, the feature maps output from DBMs provide more shape-aware local details and our proposed TransBTSV2 can generate finer segmentation results of lesion areas.

	\section{Experiments and Results}
	\label{sec:experimentsandresults}
    \subsection{Data and Evaluation Metrics}
    We evaluate our proposed method on four benchmark datasets for medical volumetric segmentation. 
    The data information and evaluation metrics of the four medical image datasets are described elaborately as following.
    
    \subsubsection{BraTS 2019 and BraTS 2020}
    
    The first 3D MRI dataset used in the experiments is provided by the Brain Tumor Segmentation (BraTS) 2019 challenge \cite{menze2014multimodal,bakas2017advancing,bakas2018identifying}. It contains 335 cases of patients for training and 125 cases for validation. Each sample is composed of four modalities of brain MRI scans. Each modality has a volume of $240\times240\times155$ which has been aligned into the same space. 
    The labels contain 4 classes: background (label 0), necrotic and non-enhancing tumor (label 1), peritumoral edema (label 2) and GD-enhancing tumor (label 4). 
    The performance on the \textbf{validation set} assessed by the online evaluation server is used to validate the effectiveness of the proposed method.
    The segmentation accuracy is measured by the Dice score and the Hausdorff distance (95\%) metrics for enhancing tumor region (ET, label 4), regions of the tumor core (TC, labels 1 and 4), and the whole tumor region (WT, labels 1,2 and 4).
    The second 3D MRI dataset is provided by the Brain Tumor Segmentation (BraTS) 2020 Challenge \cite{menze2014multimodal,bakas2017advancing,bakas2018identifying}. It consists of 369 cases for training, 125 cases for validation and 166 cases for testing. 
    Except for the number of samples in the dataset, the other information regarding these two MRI datasets are the same.
    
    \subsubsection{LiTS 2017}
    
    The third medical image dataset is provided by the Liver Tumor Segmentation (LiTS) 2017 challenge \cite{bilic2019liver}. The LiTS dataset contains 131 and 70 samples of contrast-enhanced 3D abdominal computed tomography scans with annotations of liver and liver tumors for training and testing, respectively. 
    The dataset was acquired by different scanners and protocols from six different clinical sites, with a largely varying in-plane resolution from 0.55 mm to 1.0 mm and slice spacing from 0.45 mm to 6.0 mm. 
    The number of slices ranges from 75 to 987 while the spatial size of each CT image and label is $512\times512$. The labels contain 3 classes: background (label 0), liver (label 1), liver tumor (label 2). 
    We employ the Dice per case score and Dice global score as evaluation metrics. Dice per case score refers to an average Dice score per volume and Dice global score is the Dice score evaluated by combining the whole datasets into one.
    
    \subsubsection{KiTS 2019}
    
    The fourth medical image dataset is provided by Kidney Tumor Segmentation (KiTS) 2019 Challenge \cite{heller2019kits19}. KiTS 2019 dataset provides data of multi-phase 3D CTs with high-quality annotated voxel-wise labels for 300 patients. 
    210 patients were randomly selected for the training set and the remaining 90 patients were the testing set for algorithm evaluation. 
    The spatial size of each CT image and label is $512\times512$ while the label was performed with roughly 50 annotated slices depicting the kidneys and tumors for each patient. 
    The ground truth contains 3 classes: background (label 0), kidney (label 1), kidney tumor (label 2). For evaluation of our method, the same three evaluation metrics as KiTS 2019 challenge are used. Kidney dice denotes the segmentation performance when considering both kidneys and tumors as the foreground whereas tumor dice considers everything except the tumors as background. Composite dice is the average of kidney dice and tumor dice.
    
    \subsection{Implementation Details}
    
    \noindent \textbf{Preprocessing.} 
    On BraTS 2019 and BraTS 2020 datasets, since each modality has been aligned into the same space, resampling is not needed for these two datasets. However, in order to make the gray values of each image in the training set have the same distribution, intensity normalization is necessary for input data. Specifically, for the foreground with non-zero voxel values in the medical images and corresponding labels, z-score normalization is employed within each modality.
    
    On LiTS2017 and KiTS2019 datasets, as the voxel spacings of these two CT datasets are inhomogeneous, resampling all cases to a common voxel spacing is required. Thus, we first resample all cases to the target spacing with the resulted volumetric spatial resolution for training cases. Then, since the intensity values are expected to be identical when examining the same organ on scans originating from different scanners or hospitals, each case is clipped to the appropriate organ-specific value range. At last, z-score normalization is also introduced to bring the intensity values in an expected range.
    
    \noindent \textbf{Training Details.} The proposed TransBTSV2 is implemented based on Pytorch \cite{paszke2019pytorch} and trained with NVIDIA Titan RTX GPUs. On the four benchmark datasets, TransBTSV2 is trained from scratch with a batch size of 16. On BraTS 2019 and BraTS 2020 datasets, the expansion ratio E is set to 1.5. The Adam optimizer is adopted to train the model for 6000 epochs. With warm-up strategy for 60 epochs during training, the initial learning rate is set to 0.00008 with a cosine learning rate decay schedule.
    In the training phase, random cropping, random mirror flipping and random intensity shift are applied as the data augmentation techniques.
    The softmax Dice loss is employed to train the network and $L2$ Norm is also applied for model regularization with a weight decay rate of $10^{-5}$. 
    On LiTS2017 and KiTS2019 datasets, the expansion ratio E is set to 2. Adam optimizer is also used to train our method for 6000 epochs and 12000 epochs respectively. With warm-up strategy for 80 epochs and 160 epochs respectively during training, the initial learning rate is set to 0.002 with a cosine learning rate decay schedule. In the training phase, the preprocessed data is randomly cropped to the size of $128\times128\times128$, while the padding operation may be needed to keep the number of slice consistent within each training case. 
    A number of data augmentation techniques such as rotations and scaling are applied during training, which remain consistent with \cite{isensee2021nnu}. 

    \subsection{Experimental Results}
    
    \subsubsection{Evaluation on Brain Tumor Segmentation}
    
    \begin{table*}[htbp]
    \scriptsize
        \centering
        \caption{Performance comparison on BraTS 2019 validation set. Per Case and Per Slice denote the computational costs of segmenting a 3D patient case and a single 2D slice separately.} 
        \label{tab:comparison}
        \resizebox{\linewidth}{!}
        {
        \begin{tabular}{l|c|c|c|c|c|c|c|c|c}
            \toprule[1.1pt]
            \multirow{2}{*}{Method} & \multicolumn{3}{c|}{Dice Score (\%) $\uparrow$} & \multicolumn{3}{c|}{Hausdorff Dist. (mm) $\downarrow$} & \multicolumn{2}{c|}{FLOPs (G) $\downarrow$} &\multirow{2}{*}{Params (M) $\downarrow$} \\
            \cline{2-9}
            &  ET &  WT & TC &  ET &  WT & TC & Per Case & Per Slice & \\
            \hline
            3D U-Net \cite{3dunet}  & 70.86 & 87.38 & 72.48 & 5.062 & 9.432 & 8.719 & 1,669.53 & 13.04  & 16.21 \\
            V-Net  \cite{vnet}    & 73.89 & 88.73 & 76.56 & 6.131 & 6.256 & 8.705 
            & 749.29 & 5.85  & 45.61 \\
            Attention U-Net  \cite{oktay2018attention}   & 75.96 & 88.81 & 77.20 & 5.202 & 7.756 & 8.258 & 525.11 & 4.10  & 10.88 \\
            Wang et al. \cite{wang20193d}         & 73.70 & 89.40 & 80.70 & 5.994 & 5.677 & 7.357 & - & - & -  \\
            Chen et al.  \cite{chen2019aggregating}   & 74.16 & 90.26 & 79.25 & 4.575 & 4.378 & 7.954 & - & - & - \\
            Li et al. \cite{li2019multi}           & 77.10 & 88.60 & 81.30 & 6.033 & 6.232 & 7.409 & - & - & -  \\
            Frey and Nau \cite{frey2019memory}      & 78.70 & 89.60 & 80.00 & 6.005 & 8.171 & 8.241 & - & - & - \\
            Vu et al.  \cite{vu2019tunet}   & 78.42 & 90.34 & 81.12 & 3.700 & \textbf{4.320} & 6.280 & - & - & - \\
            KiU-Net \cite{valanarasu2021kiu}        & 66.37 & 86.12 & 70.61 & 9.418 & 12.790 & 13.040  & - & - & - \\
            3D KiU-Net \cite{valanarasu2021kiu}     & 73.21 & 87.60 & 73.92 & 6.323 & 8.942 & 9.893    & \textbf{234.89} & \textbf{1.84} & \textbf{9.55} \\
            TransUNet  \cite{chen2021transunet}   & 78.17 & 89.48 & 78.91 & 4.832 & 6.667 & 7.365 & 1205.76 & 9.42  & 105.18 \\
            Swin-Unet  \cite{cao2021swin}   & 78.49 & 89.38 & 78.75 & 6.925 & 7.505 & 9.260 & 250.88 & 1.96  & 27.15 \\
            \hline
            \textbf{TransBTS} \cite{wang2021transbts}  & 78.93 & 90.00 & 81.94 & 3.736 & 5.644 & 6.049  & 333.09 & 2.60  & 32.99 \\
            \textbf{TransBTSV2 (Ours)}   & \textbf{80.24} ($\uparrow$1.31) & \textbf{90.42} ($\uparrow$0.42)& \textbf{84.87} ($\uparrow$2.93)& \textbf{3.696} ($\downarrow$0.040) & 5.432 ($\downarrow$0.212) & \textbf{5.473} ($\downarrow$0.576) & 240.66 ($\downarrow$92.43)  & 1.88 ($\downarrow$0.72)  & 15.30 ($\downarrow$17.69)  \\
            \bottomrule[1.1pt]
        \end{tabular}
        }
        
    \end{table*}

    \textbf{BraTS 2019.} To fairly evaluate the performance of our proposed TransBTSV2, we first conduct experiments on the BraTS 2019 validation set and compare TransBTSV2 with previous state-of-the-art approaches. 
    The \textbf{quantitative results} are presented in Table \ref{tab:comparison}. Our TransBTSV2 achieves the Dice scores of $80.24\%$, $90.42\%$, $84.87\%$ and Hausdorff distance of $3.696$mm, $5.432$mm, $5.473$mm on ET, WT, TC respectively, which are comparable or higher results than previous SOTA methods presented in Table \ref{tab:comparison}. In terms of the Hausdorff distance metric, a considerable improvement has been achieved for segmentation compared with previous SOTA methods. Additionally, it can be seen that TransBTSV2 obtains an absolute dominance over the Dice scores of ET and TC which is attribute to our redesigned Transformer block and the proposed DBM. Compared with 3D U-Net \cite{3dunet} that without Transformer as part of encoder, TransBTSV2 shows great superiority in both performance and efficiency with significant improvements. This clearly reveals the benefit of leveraging the redesigned Transformer block for modeling global relationships.

    \begin{figure}[htbp]
        \centering
        \includegraphics[width=0.48\textwidth]{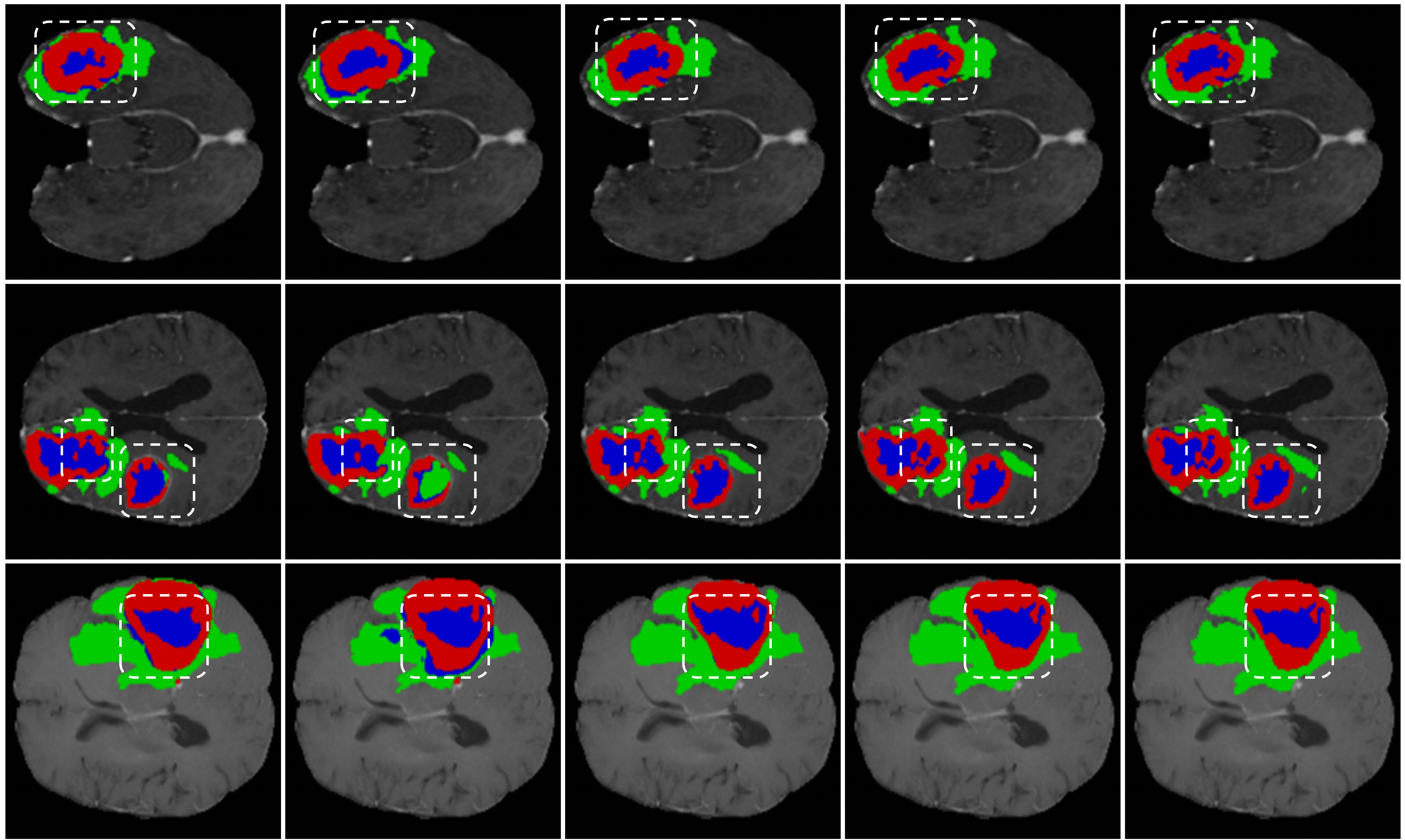}
        \begin{tabu} to 1.0\linewidth{X[1.1c] X[1.1c] X[1.0c] X[1.2c] X[1.1c]} 
            \scriptsize{3D U-Net} &  \scriptsize{VNet} &  \scriptsize{Att. U-Net} &  \scriptsize{ \textbf{TransBTSV2}} &  \scriptsize{GT} \\
        \end{tabu}
        \caption{The visual comparison of MRI brain tumor segmentation results.
        The \textcolor{blue}{blue} regions denote the enhancing tumors, the \textcolor{red}{red} regions denote the non-enhancing tumors and the \textcolor{Green3}{green} ones denote the peritumoral edema.}
        \label{fig4}
    \end{figure}   
    
    For \textbf{qualitative analysis}, we show a visual comparison of the brain tumor segmentation results of various methods including 3D U-Net\cite{3dunet} , V-Net\cite{vnet}, Attention U-Net\cite{oktay2018attention} and our TransBTSV2 in Fig.~\ref{fig4}. Since the ground truth for the validation set is not available, we conduct five-fold cross-validation evaluation on the training set for all methods. As it is shown in Fig.~\ref{fig4}, the main difference between different methods are all highlighted by white dotted boxes. It is apparent from Fig.~\ref{fig4} that TransBTSV2 can segment brain tumors more accurately and generate much better segmentation masks by modeling long-range dependencies between each volume and capturing shape-aware local information.

    \begin{table*}[htbp]
    \scriptsize
        \centering
        \caption{Performance comparison on BraTS 2020 validation set. Per Case and Per Slice denote the computational costs of segmenting a 3D patient case and a single 2D slice separately.}
        \label{tab:comparison2020}
        \resizebox{\linewidth}{!}
        {
        \setlength{\tabcolsep}{1.5mm}{
        \begin{tabular}{l|c|c|c|c|c|c|c|c|c}
            \toprule[1.1pt]
            \multirow{2}{*}{Method} & \multicolumn{3}{c|}{Dice Score (\%) $\uparrow$} & \multicolumn{3}{c|}{Hausdorff Dist. (mm) $\downarrow$} & \multicolumn{2}{c|}{FLOPs (G) $\downarrow$} &\multirow{2}{*}{Params (M) $\downarrow$}  \\
            \cline{2-9}
            & ET &  WT &  TC &  ET &  WT & TC & Per Case & Per Slice & \\
            \hline
            3D U-Net    \cite{3dunet}        & 68.76 & 84.11 & 79.06 & 50.983 & 13.366 & 13.607 & 1,669.53 & 13.04  & 16.21 \\
            Basic V-Net \cite{vnet}              & 61.79 & 84.63 & 75.26 & 47.702 & 20.407 & 12.175 & 749.29 & 5.85  & 45.61\\
            Deeper V-Net  \cite{vnet}              & 68.97 & 86.11 & 77.90 & 43.518 & 14.499 & 16.153 & - & - & - \\
            3D Residual U-Net \cite{zhang2018road}   & 71.63 & 82.46 & 76.47 & 37.422 & 12.337 & 13.105 & 407.37 & 3.18 & 9.50 \\
            Liu et al. \cite{liu2020brain}   & 76.37 & 88.23 & 80.12 & 21.390 & 6.680 & 6.490 & - & - & - \\
            Vu et al.  \cite{vu2020multi}   & 77.17 & 90.55 & 82.67 & 27.040 & 4.990 & 8.630 & - & - & - \\
            Ghaffari et al. \cite{ghaffari2020brain}   & 78.00 & 90.00 & 82.00 & - & - & -  & - & - & - \\
            Nguyen et al. \cite{nguyen2020enhancing}   & 78.43 & 89.99 & 84.22 & 24.024 & 5.681 & 9.566 & - & - & - \\
            TransUNet  \cite{chen2021transunet}   & 78.42 & 89.46 & 78.37 & 12.851 & 5.968 & 12.840 & 1205.76 & 9.42  & 105.18 \\
            Swin-Unet  \cite{cao2021swin}   & 78.95 & 89.34 & 77.60 & \textbf{11.005} & 7.855 & 14.594 & 250.88 & 1.96  & 27.15 \\
            \hline
            \textbf{TransBTS} \cite{wang2021transbts}   & 78.73 & 90.09 & 81.73 & 17.947 & 4.964 & 9.769  & 333.09 & 2.60  & 32.99  \\
            \textbf{TransBTSV2 (Ours)}          & \textbf{79.63} ($\uparrow$0.90) & \textbf{90.56} ($\uparrow$0.47) & \textbf{84.50} ($\uparrow$2.77) & 12.522 ($\downarrow$5.425) & \textbf{4.272} ($\downarrow$0.692) & \textbf{5.560} ($\downarrow$4.209)  & \textbf{240.66} ($\downarrow$92.43)  & \textbf{1.88} ($\downarrow$0.72)  & \textbf{15.30} ($\downarrow$17.69) \\
            \bottomrule[1.1pt]
        \end{tabular}
        }
        }
    \end{table*}

    \noindent \textbf{BraTS 2020.} We also evaluate TransBTSV2 on BraTS 2020 validation set and the results are reported in Table \ref{tab:comparison2020}. 
    TransBTSV2 achieves Dice scores of $79.63\%$, $90.56\%$, $84.50\%$ and Hausdorff distance of 12.522mm, 4.272mm, 5.560mm on ET, WT, TC respectively. Similar to the evaluation on BraTS 2019 validation set, our TransBTSV2 again demonstrate the absolution superiority on both the the Dice scores and Hausdorff distance on ET and TC. Compared with 3D U-Net\cite{3dunet}, V-Net \cite{vnet} and Residual U-Net \cite{zhang2018road}, TransBTSV2 shows great advantages in both metrics with prominent improvements. Moreover, TransBTSV2 reaches competitive or better performance than previous SOTA approaches on BraTS 2020 validation set, demonstrating the potential of our well-designed hybrid architecture. To be mentioned, in comparison with our original TransBTS \cite{wang2021transbts} and recently proposed Transformer-based methods (e.g. TransUNet \cite{chen2021transunet} and Swin-Unet \cite{cao2021swin}), great advancements have been made by TransBTSV2 in terms of both efficiency and accuracy on BraTS 2019 and BraTS 2020 datasets.

    \subsubsection{Evaluation on Liver Tumor Segmentation}

    \begin{table}[htbp]
    \scriptsize
        \centering
        \caption{Performance comparison on LiTS 2017 testing set. Per Case and Per Slice denote the computational costs of segmenting a 3D patient case and a single 2D slice separately, P refers to the pre-trained model.} 
        \label{tab:comparisonlits2017}
        \resizebox{\linewidth}{!}
        {
        \setlength{\tabcolsep}{0.6mm}{
        \begin{tabular}{l|c|c|c|c|c|c|c}
            \toprule[1.1pt]
            \multirow{2}{*}{Method} & \multicolumn{2}{c|}{Dice Per Case (\%) $\uparrow$} & \multicolumn{2}{c|}{Dice Global (\%) $\uparrow$} & \multicolumn{2}{c|}{FLOPs (G) $\downarrow$} &\multirow{2}{*}{Params (M) $\downarrow$} \\
            \cline{2-7}
            & Lesion &  Liver &  Lesion &  Liver & Per Case & Per Slice & \\
            \hline
            Han \cite{han2017automatic}           & 67.00 & - & - & -  & - & - & - \\
            Chlebus et al. \cite{chlebus2017neural}  & 65.00 & - & - & - & - & - & - \\
            I3D  \cite{carreira2017quo}   & 62.40 & 95.70 & 77.60 & 96.00  & - & - & - \\
            I3D w/ P  \cite{carreira2017quo}   & 66.60 & 95.60 & 79.90 & 96.20  & - & - & - \\
            Vorontsov et al. \cite{vorontsov2018liver}         & 65.00 & - & - & -  & - & - & - \\
            3D DenseUNet w/o P \cite{li2018h}   & 59.40 & 93.60 & 78.80 & 92.90  & - & - & - \\
            2D DenseUNet w/o P  \cite{li2018h}   & 67.70 & 94.70 & 80.10 & 94.70  & - & - & -\\
            2D DenseNet w/ P  \cite{li2018h}   & 68.30 & 95.30 & 81.80 & 95.90  & - & - & - \\
            2D DenseUNet w/ P  \cite{li2018h}   & 70.20 & 95.80 & 82.10 & 96.30  & - & - & - \\
            TransUNet  \cite{chen2021transunet}   & 61.70 & 95.40 & 77.40 & 95.60  & 1200.64 & 9.38 & 105.17\\
            Swin-Unet  \cite{cao2021swin}   & - & 92.70 & 67.60 & 91.60 & 249.60 & 1.95 & 27.15 \\
            \hline
            \textbf{TransBTS} \cite{wang2021transbts}  & 70.30  &  96.00 & 81.50  & 96.40  
            & 330.37  & 2.58  & 32.79 \\
            \textbf{TransBTSV2 (Ours)}  & \textbf{71.20} ($\uparrow$0.90) & \textbf{96.20} ($\uparrow$0.20)  & \textbf{83.10} ($\uparrow$1.60) & \textbf{96.60} ($\uparrow$0.20) 
            & \textbf{237.94} ($\downarrow$92.43)  & \textbf{1.86} ($\downarrow$0.72)  & \textbf{15.30} ($\downarrow$17.49) \\
            \bottomrule[1.1pt]
        \end{tabular}
        }
        }
    \end{table}

    To evaluate the performance and generalization ability of our TransBTSV2, experiments are also conducted on LiTS 2017 dataset. The performance comparison of evaluation metrics are shown in Table \ref{tab:comparisonlits2017}. It can be found that our TransBTSV2 once more achieves competitive or better performance across a wide range of metrics compared with previous SOTA approaches. Especially for the Dice metrics of lesion areas, a considerable improvement is achieved by our TransBTSV2, benefiting from the exploitation of DBM and modified Transformer block. By fully taking advantage of the volumetric spatial information in medical images, our TransBTSV2 can more accurately segment both the liver and liver tumors in a highly efficient manner compared with recent Transformer-based methods (e.g. TransUNet \cite{chen2021transunet} and Swin-Unet \cite{cao2021swin}).

    \begin{figure}[htbp]
        \centering
        \includegraphics[width=0.48\textwidth]{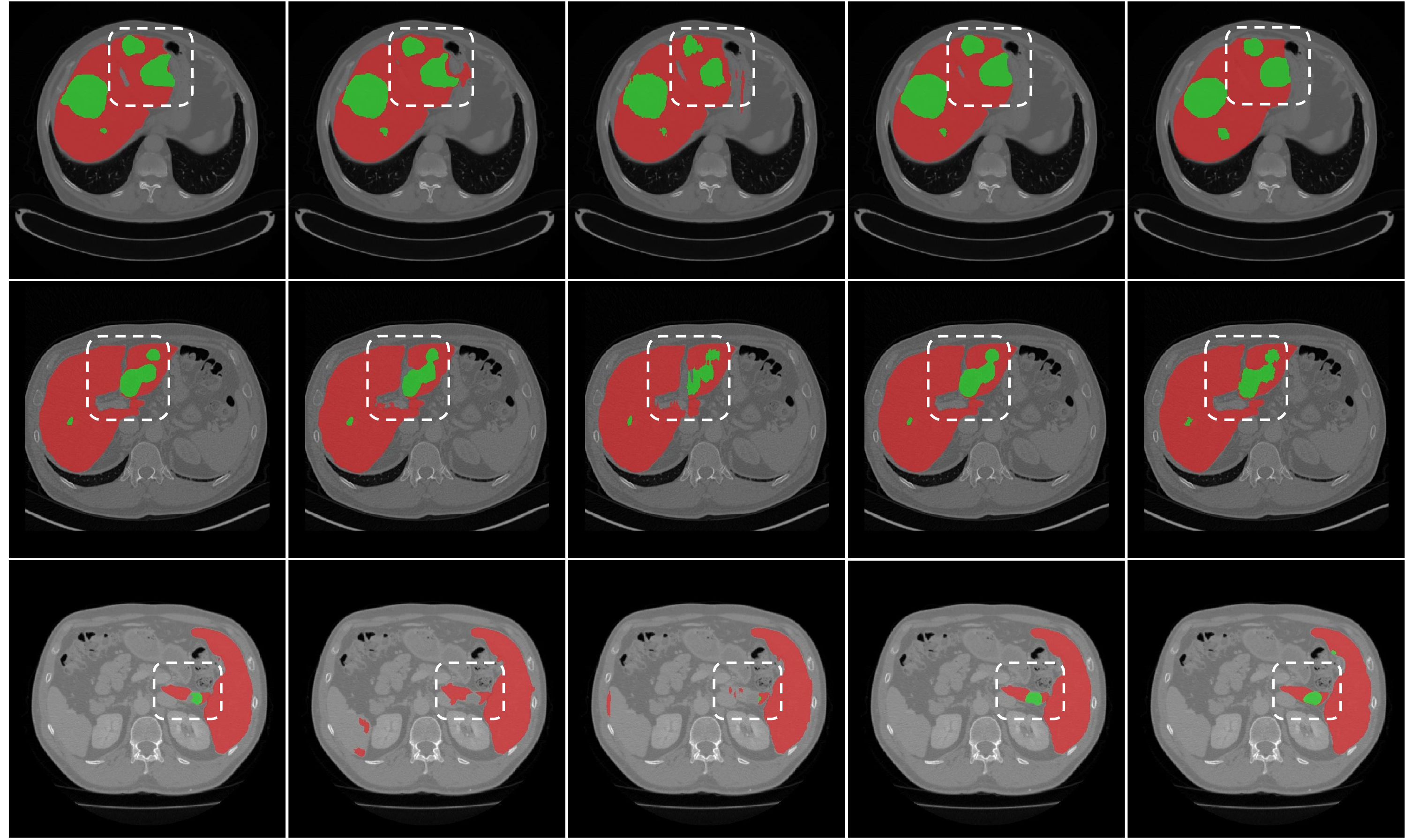}
        \begin{tabu} to 1.0\linewidth{X[1.1c] X[1.1c] X[1.0c] X[1.2c] X[1.1c]} 
            \scriptsize{3D U-Net} &  \scriptsize{VNet} &  \scriptsize{Att. U-Net} &  \scriptsize{\textbf{TransBTSV2}} &  \scriptsize{GT} \\
        \end{tabu}
        \caption{The visual comparison of CT liver tumor segmentation results. The \textcolor{red}{red} regions denote the liver and the \textcolor{Green3}{green} regions denote the tumors.}
        \label{figlits}
    \end{figure}

    For further visual comparison, the segmentation results of different methods are illustrated in Fig.~\ref{figlits}. It is evident that the segmentation results of our method are much more close to the ground truth compared with U-Net \cite{3dunet}, V-Net \cite{vnet}, and Attention U-Net \cite{oktay2018attention}. With the merits of both Transformers and convolutions, TransBTSV2 can segment liver tumors more precisely and generate finer segmentation results.

    \subsubsection{Evaluation on Kidney Tumor Segmentation}
    
    \begin{table}[htbp]
    \scriptsize
        \centering
        \caption{Performance comparison on KiTS 2019 testing set. Per Case and Per Slice denote the computational costs of segmenting a 3D patient case and a single 2D slice separately.} 
        \label{tab:comparisonkits2019}
        \resizebox{\linewidth}{!}
        {
        \setlength{\tabcolsep}{0.2mm}{
        \begin{tabular}{l|c|c|c|c|c|c}
            \toprule[1.1pt]
            \multirow{2}{*}{Method} & \multirow{2}{*}{Kidney Dice (\%) $\uparrow$} & \multirow{2}{*}{Tumor Dice (\%) $\uparrow$} 
            & \multirow{2}{*}{Composite Dice (\%) $\uparrow$} & \multicolumn{2}{c|}{FLOPs (G) $\downarrow$} &\multirow{2}{*}{Params (M) $\downarrow$} \\
            \cline{5-6}
             &   &   &   & Per Case & Per Slice & \\
            \hline
            U-Net \cite{unet}                & 95.15 & 82.45 & 88.80 & 1457.92 & 11.39 & 23.38 \\
            V-Net  \cite{vnet}               & 93.70 & 80.72 & 87.21 & 736.70 & 5.76 & 45.60 \\
            Myronenko \cite{myronenko20183d}   & 95.30 & 82.35 & 88.83 & - & - & - \\
            BiSC-UNet  \cite{wang2019bisc}              & 95.40 & 74.10 & 84.75 & - & - & - \\
            Ma \cite{ma2019solution}               & 97.34 & 82.54 & 89.94  & - & - & - \\
            Li \cite{li2019fully}                & 97.17 & 81.61 & 89.39  & - & - & - \\
            Chen \cite{chen2019segmentation}            & 97.01 & 81.40 & 89.20  & - & - & - \\
            SERU \cite{xie2020seru}         & 96.80 & 74.30 & 85.55 & - & - & - \\
            TransUNet  \cite{chen2021transunet}   & 95.43 & 66.08 & 80.75  & 1200.64 & 9.38 & 105.17 \\
            Swin-Unet  \cite{cao2021swin}   & 93.77 & - & -  & 249.60 & 1.95 & 27.15  \\        
            \hline
            \textbf{TransBTS} \cite{wang2021transbts}  & 96.78 & 81.42 & 89.10  & 330.37  & 2.58  & 32.79 \\
            \textbf{TransBTSV2 (Ours)}     & \textbf{97.37} ($\uparrow$0.59) &  \textbf{83.69} ($\uparrow$2.27) & \textbf{90.53} ($\uparrow$1.43)  & \textbf{237.94} ($\downarrow$92.43)  & \textbf{1.86} ($\downarrow$0.72)  & \textbf{15.30} ($\downarrow$17.49)  \\
            \bottomrule[1.1pt]
        \end{tabular}
        }
        }
        
    \end{table}
    
    To further validate the effectiveness and robustness of our method, we conduct experiments on the KiTS 2019 dataset, which provides images and labels of kidneys and lesions with high varieties and complexities. Table \ref{tab:comparisonkits2019} shows the performance comparison of tumor and kidney segmentation between our method and previous SOTA methods. 
    According to Table \ref{tab:comparisonkits2019}, TransBTSV2 significantly enhances the segmentation accuracy for both kidneys and tumors, demonstrating the effectiveness and robustness of our well-designed hybrid architecture. 
    In contrast with original TransBTS \cite{wang2021transbts}, significant improvements ($\uparrow$2.27\% and $\uparrow$1.43\% respectively) have been made on the metrics of tumor dice score and composite dice score. As a hybrid architecture of 3D CNN and Transformer, our TransBTSV2 can jointly exploit the volumetric spatial information (i.e. spatial and slice correlations) with the merits of both convolution operation and self-attention mechanism. Compared with recent Transformer-based methods (e.g. TransUNet \cite{chen2021transunet} and Swin-Unet \cite{cao2021swin}), significant improvements have been made by our TransBTSV2 on both model performance and efficiency, which is attributed to our proposed insight to redesign Transformer block and the introduced Deformable Bottleneck Module.
    
\begin{figure}[htbp]
        \centering
        \includegraphics[width=0.48\textwidth]{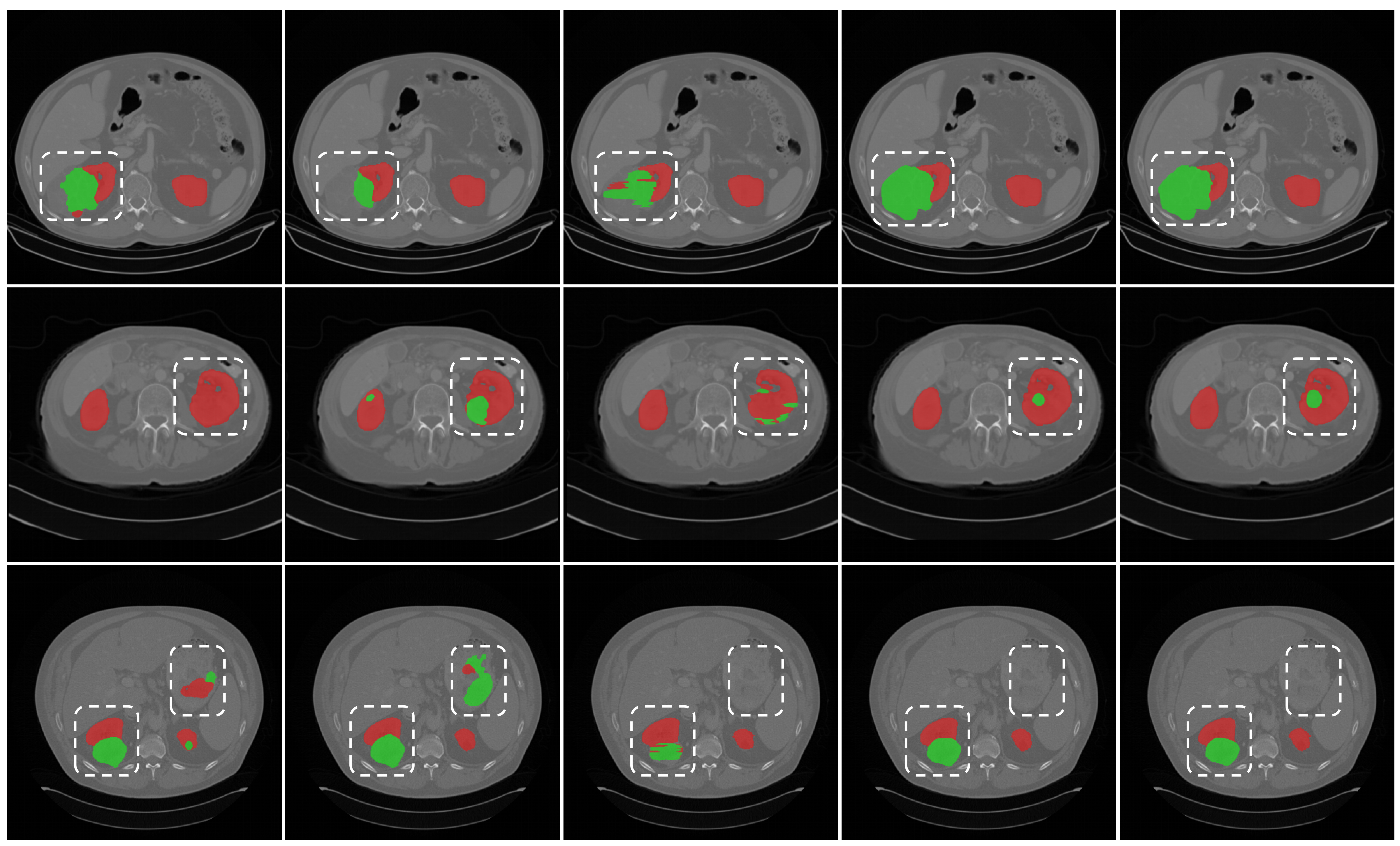}
        \begin{tabu} to 1.0\linewidth{X[1.1c] X[1.1c] X[1.0c] X[1.2c] X[1.1c]} 
            \scriptsize{3D U-Net} &  \scriptsize{VNet} &  \scriptsize{Att. U-Net} &  \scriptsize{\textbf{TransBTSV2}} &  \scriptsize{GT} \\
        \end{tabu}
        \caption{The visual comparison of CT kidney tumor segmentation results. The \textcolor{red}{red} regions denote the kidneys and the \textcolor{Green3}{green} regions denote the tumors.}
        \label{figkits2D}
    \end{figure}

    Furthermore, to comprehensively analyze the quality of segmentation results, the 2D slices of kidneys and tumors are shown in Fig.~\ref{figkits2D}. It is clear that with the advantages of exploiting Transformer and the proposed DBM to model long-range dependencies and deformation of irregular shapes, both small-scale tumors and relatively large-scale kidneys can be well segmented by our proposed TransBTSV2.

    \subsection{Model Complexity}
    
    In the MICCAI version \cite{wang2021transbts}, our TransBTS is a moderate-size model with 32.99M parameters and 333.09G FLOPs. 
    With the improved architecture design proposed in this paper, TransBTSV2 only has 15.30M parameters and 240.66G FLOPs while achieving Dice scores of $80.24\%$, $90.42\%$, $84.87\%$ and Hausdorff distance of 3.696mm, 5.432mm, 5.473mm on ET, WT, TC on BraTS2019 validation set. 
    It is worth noting that compared with our original TransBTS \cite{wang2021transbts}, pursuing wider instead of deeper Transformer leads to greatly reduced complexity ($\downarrow\textbf{53.62\%}$ in parameters and $\downarrow\textbf{27.75\%}$ in FLOPs) but significantly improves model performance in the meantime. Compared with 3D U-Net \cite{3dunet} which has 16.21M parameters and 1669.53G FLOPs, our TransBTSV2 shows great superiority in terms of computational complexity as well as model performance.

    \subsection{Ablation Study}
    
    We conduct extensive ablation experiments to verify the effectiveness of TransBTSV2 and justify the rationality of its design based on evaluations on BraTS 2019 dataset. Since there is no testing set provided by the challenge organizers, the ablation experiments are performed on official \textbf{validation set} for fair comparison. 
    (1) We investigate the effectiveness of each introduced component in our TransBTSV2. 
    (2) We make a thorough inquiry of different designs for spatial-slice attention in Transformer. 
    (3) We further analyze the influence of different positions of our well-designed DBM.
    (4) Last but not least, we probe into the impact of different designs for pursuing the width or depth of Transformer blocks.

    \begin{table}[htpb]
    \scriptsize
        \centering
        \caption{Ablation study for each introduced component on the BraTS 2019 validation set. B, TR, FEM, DBM, QK$\uparrow$ refers to baseline, Transformer, feature expansion module, Deformable Bottleneck Module, the proposed FW-MHSA in redesigned Transformer block respectively.}
        \label{tab:ablation}
        \setlength{\tabcolsep}{1.4mm}{
        \begin{tabular}{l|c|c|c|c|c}
            \toprule[1.1pt]
            \multirow{2}{*}{Method} & \multicolumn{3}{c|}{Dice Score (\%) $\uparrow$} & \multirow{2}{*}{Params(M)} & \multirow{2}{*}{FLOPs(G)}\\
            \cline{2-4}
            &  ET &  WT &  TC & \\
            \hline
            B  & 77.03 & 89.72 & 78.28  & 4.76  & 158.44\\
            B + TR      & 77.86 & 90.10 & 82.29  & 5.48 & 163.55\\
            B + TR + FEM     & 80.30  & 90.30  & 83.84  & 14.88  & 208.47\\
            B + TR + FEM + DBM      & \textbf{80.80} & 90.34 & 84.00 & 15.03 & 235.22\\
            B + TR + FEM + DBM + QK$\uparrow$     & 80.24 & \textbf{90.42} & \textbf{84.87} & 15.30 & 240.66\\
            \bottomrule[1.1pt]
        \end{tabular}
        }
    \end{table}
    
    \textbf{Ablation Study for Each Component of our TransBTSV2.} We first investigate the effectiveness of each introduced component in our TransBTSV2. Noticeably, a modified 3D UNet with lightweight structure and inserted pre-activated residual blocks is employed as our optimal baseline. As illustrated in Table~\ref{tab:ablation}, compared with our baseline, the hybrid CNN-Transformer architecture shows great superiority on the metric of Dice scores, leading to considerable performance improvements ($\uparrow$0.83\%, $\uparrow$0.38\% and $\uparrow$4.01\% on ET, WT and TC respectively). This clearly reveals the benefit of leveraging Transformer to model the global relationships. Moreover, employing the proposed DBM makes our TransBTSV2 more powerful to model irregular-shape deformation of lesion regions, while the introduced feature expansion module and the proposed insight of pursing an inverted bottleneck alike architecture (i.e. pursue Transformer width instead of depth) both empower TransBTSV2 to possess richer feature representations. 
    To be specific, the introduction of the FEM brings remarkable gain to all the Dice scores ($\uparrow$2.44\%, $\uparrow$0.20\%, $\uparrow$1.55\% separately) and the proposed insight also leads to a great advancement on the Dice score of WT and TC, jointly enabling our proposed TransBTSV2 to better solve the inherent problems of medical image segmentation.

    \begin{table}[htpb]
    \scriptsize
        \centering
        \caption{Ablation study on different designs for spatial-slice self-attention on the BraTS 2019 validation set.} 
        \label{tab:spatial-slice attention}
        {
        \begin{tabular}{l|c|c|c|c|c}
            \toprule[1.1pt]
            \multirow{2}{*}{Method} & \multicolumn{3}{c|}{Dice Score (\%) $\uparrow$} & \multirow{2}{*}{Params(M)} & \multirow{2}{*}{FLOPs(G)} \\
            \cline{2-4}
                & \bfseries ET & \bfseries WT & \bfseries TC &  &  \\
            \hline
            spatial-only    & 79.27  & 89.88 & 82.97 &  \textbf{13.17}  &  \textbf{192.38}  \\
            split-cascaded  & 79.09  & \textbf{90.68} & 83.27 & 16.33  & 205.36 \\
            split-parallel  & 79.33  & 90.50 & 83.16 & 16.33  & 205.36 \\
            \textbf{joint}   & \textbf{80.30}  & 90.30  & \textbf{83.84}  & 14.88  & 208.47 \\        
            \bottomrule[1.1pt]
        \end{tabular}
        }
        
    \end{table}

    \textbf{Ablation Study for Different Designs of Spatial-Slice Self-Attention in Transformer.} We also make a thorough inquiry on different designs for spatial-slice self-attention in Transformer. The joint version collectively computes spatial-slice self-attention at the same time,  the split version decouples joint attention into spatial attention and slice attention respectively (i.e. spatial Transformer and slice Transformer), and spatial-only denotes only computing self-attention at spatial dimension. Cascaded and parallel represent the relative positions of the two kinds of decoupled Transformer (i.e. spatial Transformer and slice Transformer) mentioned above. As shown in Table~\ref{tab:spatial-slice attention}, the joint version, computing self-attention at spatial and slice dimension simultaneously, achieves a better trade-off between model performance and model complexity than other designs (split version and spatial-only version).
    A considerable improvement is achieved by the joint version especially on the Dice scores of ET and TC. Although the split version and spatial-only version can reduce much computational costs, due to the limited number of tokens (i.e. sequence length) in a single dimension (i.e. spatial or slice),
    they cannot effectively model explicit long-range dependencies among volumetric spatial features, leading to a sharp decline in model performance.
    
    \begin{table}[htpb]
    \scriptsize
        \centering
        \caption{Ablation study on different positions of our DBM on the BraTS 2019 validation set.} 
        \label{tab:positions of DBM}
        {
        \begin{tabular}{l|c|c|c|c|c|c}
            \toprule[1.1pt]
            \multirow{2}{*}{Method} & \multicolumn{3}{c|}{Dice Score (\%) $\uparrow$} & \multicolumn{3}{c}{Hausdorff Dist. (mm) $\downarrow$} \\
            \cline{2-7}
            &  ET &  WT &  TC &  ET &  WT &  TC\\
            \hline
            w/o DBM  & 80.30  & 90.30  & 83.84  & 3.968  & 4.435  & 5.699 \\
            parallel with MHSA  & 79.81  & 90.26  & 83.89  & 3.640  & 5.339  & 5.549 \\
            replace FFN  & 79.38  & 90.14  & 83.91  & 3.908  & \textbf{4.361}  & 6.126 \\
            \textbf{on skip-connections}  & \textbf{80.80} & \textbf{90.34} & \textbf{84.00} & \textbf{3.486} & 4.586 & \textbf{5.272}\\
            \bottomrule[1.1pt]
        \end{tabular}
        }
    \end{table}

    \textbf{Ablation Study for Different Positions of Deformable Bottleneck Module.} In addition, we further analyze the influence of different positions of the DBMs. The experimental results are reported in Table~\ref{tab:positions of DBM}. Compared with the architecture without DBM, employing the proposed module on skip-connections brings only 0.15M parameters but considerable improvements to both the Dice scores and Hausdorff distance of ET and TC, which validates the effectiveness and efficiency of our DBM to capture more shape-aware local details. 
    Moreover, we also attempt to integrate the DBM with Transformer (i.e. parallel with multi-head self-attention or replace the FFN). As shown in Table~\ref{tab:positions of DBM}, the resulting networks can not perform as well as before. 
    We speculate that the local features generated by the inserted DBM inside Transformer may interfere with the long-range feature modeling of the original Transformer, leading to worse performance. 
    
    \begin{table}[htpb]
    \scriptsize
        \centering
        \caption{Ablation study on different designs for pursuing width or depth of Transformer blocks on the BraTS 2019 validation set. V1, V2 refers to our TransBTS in the MICCAI version and the proposed TransBTSV2 in this work respectively, while D-N, S-W denotes the architecture design of deep and narrow, the architecture design of shallow and wide respectively.} 
        \label{tab:width or depth}
        \resizebox{\linewidth}{!}
        {
        \begin{tabular}{l|c|c|c|c|c}
    
            \toprule[1.1pt]
            \multirow{2}{*}{Method} & \multicolumn{3}{c|}{Dice Score (\%) $\uparrow$} & \multirow{2}{*}{Params(M)} & \multirow{2}{*}{FLOPs(G)} \\
            \cline{2-4}
                &  ET &  WT &  TC &  &  \\
            \hline
            D-N V2    &  79.50   &  90.06  &  82.26  &  15.37  &  292.47  \\
            \textbf{S-W V2}   & \textbf{80.24} ($\uparrow$0.74) & \textbf{90.42} ($\uparrow$0.36)  & \textbf{84.87} ($\uparrow$2.61)  & \textbf{15.30} ($\downarrow$0.07) & \textbf{240.66} ($\downarrow$51.81) \\
            \hline
            V1 (D-N) & 78.93  & 90.00 & 81.94 &  32.99  &  333.09  \\
            \textbf{V2 (S-W)} & \textbf{80.24} ($\uparrow$1.31) & \textbf{90.42} ($\uparrow$0.42)  & \textbf{84.87} ($\uparrow$2.93) & \textbf{15.30} ($\downarrow$17.69) & \textbf{240.66} ($\downarrow$92.43) \\
            \bottomrule[1.1pt]
        \end{tabular}
        }
        
    \end{table}
    
    \textbf{Ablation Study for Different Designs for Pursuing Width or Depth of Transformer Blocks.} At last, we probe into the impact of different designs for pursuing the width or depth of Transformer blocks. We first conduct experiments to verify the effectiveness of pursuing width instead of depth under the setting of fixed model parameters, then the model performance and computational costs of our proposed TransBTSV2 and TransBTS are also thoroughly compared to further validate our proposed insight. The quantitative results are presented in Table~\ref{tab:width or depth}. For a fair comparison, we first redesign our TransBTSV2 into a deeper and narrower structure (i.e. number of stacked Transformer blocks $L=8$ and feature embedding dimension $d=256$) with the approximately identical model size. It can be clearly seen that going wider instead of deeper enables our TransBTSV2 to greatly advance both model performance and computational costs with the same model parameters. Besides, it also proves that pursing model depth (i.e. repeatedly stacking Transformer blocks) is not always the optimal choice for architectural design, increasing the model width in a well-designed manner can similarly help the networks to possess stronger representation capabilities. In comparison with our original TransBTS \cite{wang2021transbts} that represents the conventional design of pursuing depth (i.e. repeatedly stacking Transformer blocks), significant improvements (Dice Scores of $\uparrow1.31\%$ and $\uparrow2.93\%$ on ET and TC separately) have been made by TransBTSV2 with the proposed insight to redesign the internal structure of Transformer block.

    \subsection{Comparison between TransBTS and TransBTSV2}
    
    In order to make a thorough comparison between our proposed TransBTSV2 and the previous TransBTS in the MICCAI version \cite{wang2021transbts}, we further present comprehensive analysis of these two models in terms of the other three aspects (i.e. model optimization, feature representation and prediction confidence) to prove the powerful potential of our TransBTSV2.
    
    \begin{figure}[htbp]
        \centering
        \includegraphics[width=0.42\textwidth]{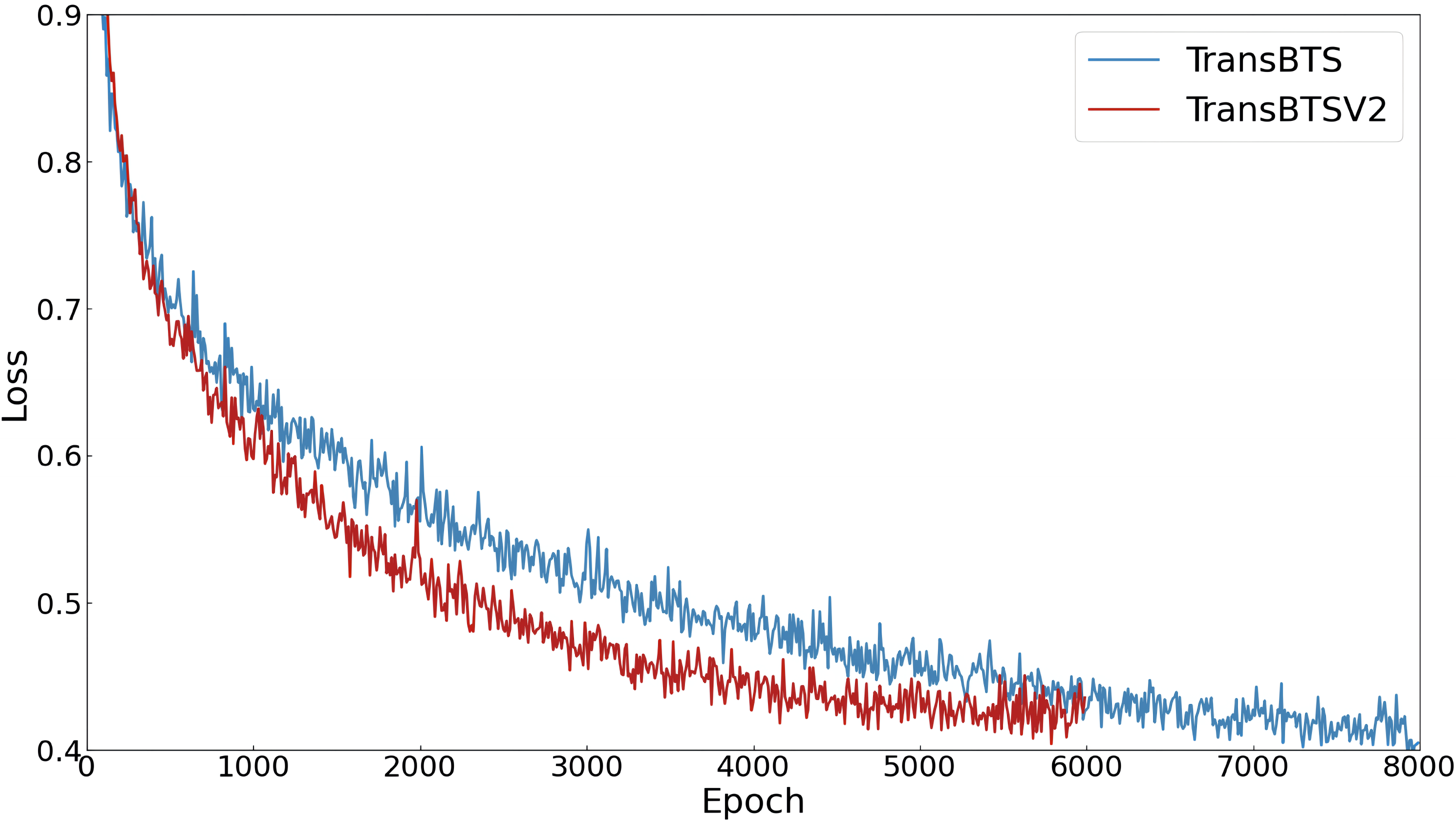}
        \caption{The visual comparison of the loss curve between the original TransBTS and our proposed TransBTSV2 during training.}
        \label{fig8}
    \end{figure}
    
    \textbf{Analysis of Model Optimization.} We first compare our proposed TransBTSV2 with previous TransBTS \cite{wang2021transbts} on model optimization. Under the same experimental setting, we record the main loss curve of TransBTS \cite{wang2021transbts} and TransBTSV2 during the training process. As illustrated in Fig.~\ref{fig8}, the red line that represents the main loss curve of our TransBTSV2 fluctuates sharply in the early stages during training and quickly converges to the ideal state for model optimization. In contrast, the fluctuation of the blue line which denotes the loss curve of TransBTS \cite{wang2021transbts} is relatively slow and the blue line finally converges to a larger value. It can be clearly seen that, with the advantage of the proposed insight to redesign our TransBTSV2 to a shallower but wider architecture (i.e. more parallel instead of deeper), TransBTSV2 is easier to be optimized and can converge faster to the final stable state in comparison with original TransBTS \cite{wang2021transbts}. Besides, it is worth emphasizing that, without any pre-training, our TransBTSV2 can save much less training time costs (i.e. $\downarrow$2000 epochs) but greatly advances the model performance compared with TransBTS \cite{wang2021transbts}.

    \begin{figure}[htbp]
        \centering
        \includegraphics[width=0.49\textwidth]{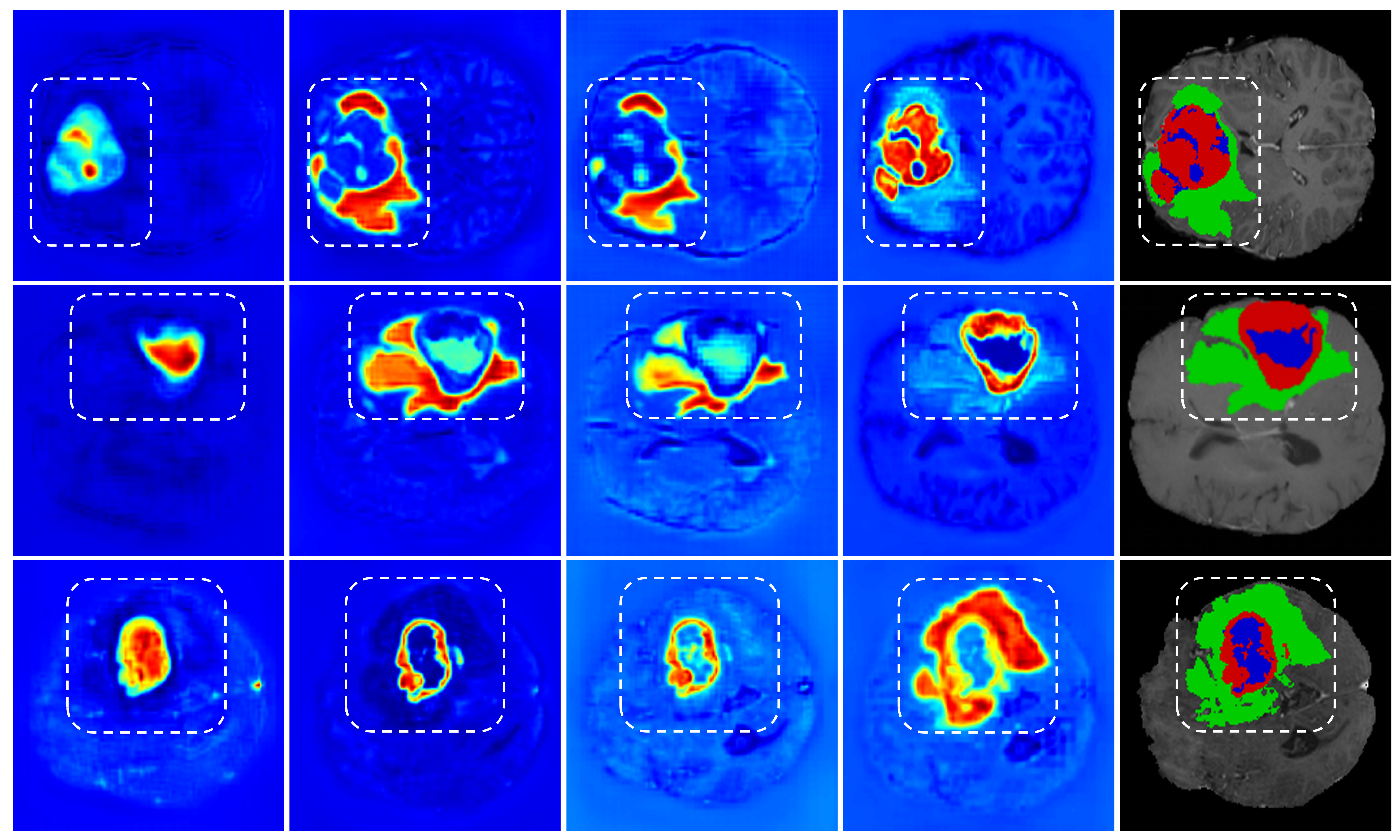}
        \begin{tabu} to 1.0\linewidth{X[1.0c] X[1.0c] X[1.0c] X[1.0c] X[1.0c]}
            \fontsize{5.5pt}{\baselineskip}\selectfont TransBTS-C1&  
            \fontsize{5.5pt}{\baselineskip}\selectfont TransBTSV2-C1 & 
            \fontsize{5.5pt}{\baselineskip}\selectfont TransBTS-C2 &  
            \fontsize{5.5pt}{\baselineskip}\selectfont TransBTSV2-C2 & 
            \fontsize{5.5pt}{\baselineskip}\selectfont GT \\
        \end{tabu}
        \caption{The visualization of feature maps output from top of the encoder of TransBTS and the proposed TransBTSV2. In the feature maps, \textcolor{red}{redder} point denotes a larger value, \textcolor{blue}{bluer} point denotes a smaller value and C1-C2 denote the corresponding channel number of the image slice . In the ground truth, the \textcolor{blue}{blue} regions denote the enhancing tumors, the \textcolor{red}{red} regions denote the non-enhancing tumors and the \textcolor{Green3}{green} ones denote the peritumoral edema.}
        \label{fig9}
    \end{figure}

    \textbf{Analysis of Feature Representation.} To further investigate the learned feature representation of TransBTS \cite{wang2021transbts} and our proposed TransBTSV2, we visualize the feature maps output from the last stage of decoder for qualitative analysis in Fig.~\ref{fig9}. Specifically, two prominent channels of the output feature maps are selectively visualized for each image slice to make a fair comparison of learned feature representations. The visualization can be reckoned as heatmaps, in which red color indicates a larger pixel value and blue color indicates a relatively smaller value. 
    In comparison with TransBTS \cite{wang2021transbts}, the values in feature maps of TransBTSV2 have a higher degree of similarity for both the foreground and background pixels. Besides, compared with TransBTS \cite{wang2021transbts}, the feature maps output from our proposed TransBTSV2 have a clearer contour and inter-class contrast in the lesion areas. With the merits of both the proposed DBM and the redesigned Transformer block to model shape-aware local information and long-range dependencies, our TransBTSV2 can obviously more focus on tumor regions and capture more representative features. 
    Furthermore, although the lesion areas in the corresponding slice is highly irregular in shape and size, it can be clearly noticed that the learned features of our proposed TransBTS is much more closer to the ground truth, which fully proves the vital importance of our DBM and the proposed redesign insight to achieve the accurate segmentation of tumor regions.
    
    \begin{figure}[htbp]
        \centering
        \includegraphics[width=0.48\textwidth]{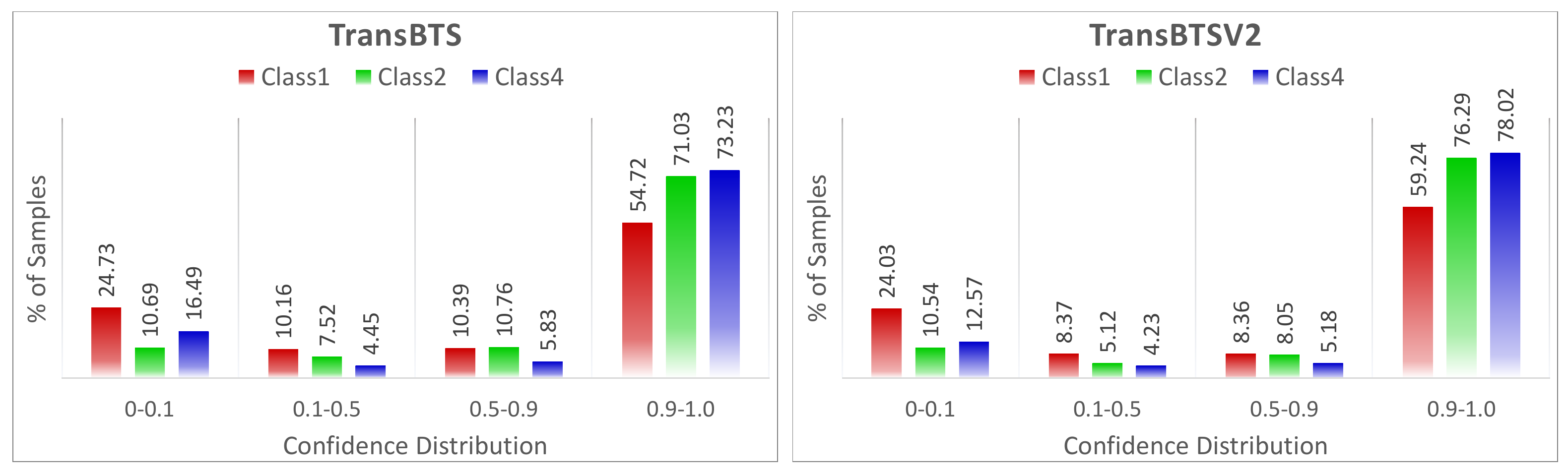}
        \caption{The confidence distribution comparison of model predictions based on the five-fold cross-validation evaluation on the BraTS 2019 training set. The \textcolor{blue}{blue} color denotes the enhancing tumors (class 4), the \textcolor{red}{red} color denotes the non-enhancing tumors (class 1) and the \textcolor{Green3}{green} color denotes the peritumoral edema (class 2).}
        \label{fig10}
    \end{figure}
    
    \textbf{Analysis of Confidence Distribution of Model Prediction.} Since the confidence of model prediction is also a vital aspect for the evaluation of network robustness, we additionally compare the prediction confidence distribution of TransBTS \cite{wang2021transbts} and TransBTSV2 based on the five-fold cross-validation evaluation on the BraTS 2019 training set. In particular, the prediction probability of each model is divided into ten intervals from 0 to 1 and the proportion of each prediction interval is depicted in Fig.~\ref{fig10}. Noticeably, the middle intervals (i.e. [0.1, 0.9]) has been merged into two intervals, [0.1, 0.5] and [0.5, 0.9], because the prediction probability is mainly distributed in the intervals of [0, 0.1] and [0.9, 1.0]. As presented in Fig.~\ref{fig10}, the prediction probabilities of our proposed TransBTSV2 more frequently fall within the intervals of [0.9, 1,0] compared with TransBTS \cite{wang2021transbts}, which indicates a higher degree of prediction confidence with improvements of $\uparrow$4.52$\%$, $\uparrow$5.26$\%$ and $\uparrow$4.79$\%$ on class 1, class 2 and class 4 respectively. In addition, the decrease ($\downarrow$3.82$\%$, $\downarrow$5.11$\%$ and $\downarrow$0.87$\%$ on class 1, class 2 and class 4 separately) in the prediction proportion of the middle interval (i.e. [0.1, 0.9]) further demonstrates that our TransBTSV2 can discard the vacillating predictions and make firmer prediction in comparison with original TransBTS \cite{wang2021transbts}.

	\section{Conclusion}
	\label{sec:conclusion}
    In this paper, we present the further attempt towards better and more efficient volumetric segmentation of medical images. Specifically, we propose a novel segmentation framework incorporating Transformer in 3D CNN for volumetric segmentation of medical images. As a hybrid CNN-Transformer architecture, the resulting architecture TransBTSV2, not only inherits the advantage of 3D CNN for modeling local context information, but also enjoys the superior long-range dependencies modeling capability of Transformer, 
    To be emphasized, TransBTSV2 can perform accurate segmentation without any pre-training, serving as a strong and efficient 3D baseline for volumetric segmentation of general medical images. 
    Experimental results on four medical image datasets demonstrate that, with the proposed insight to redesign the internal structure of Transformer block and the introduced Deformable Bottleneck Module to capture shape-aware local details, the proposed TransBTSV2 possess great generalization capability and achieves better or comparable performance compared to previous state-of-the-art methods. 
	
	\section*{Acknowledgment}
    This work was supported by the Fundamental Research Funds for the China Central Universities of USTB (FRF-DF-19-002), Scientific and Technological Innovation Foundation of Shunde Graduate School, USTB (BK20BE014).
	
	\ifCLASSOPTIONcaptionsoff
	\newpage
	\fi

	\bibliographystyle{IEEEtran}
	\bibliography{reference}
	
	%

\end{document}